\documentclass[twocolumn,superscriptaddress,floatfix,preprintnumbers,prl ,nofootinbib,hyperref,longbibliography]{revtex4-1}
\usepackage{enumerate}
\usepackage{mathrsfs,amsmath,amssymb}
\usepackage{natbib}
\usepackage{graphicx}
\usepackage{cancel}
\usepackage[normalem]{ulem}
\usepackage[colorlinks, linkcolor=red, anchorcolor=blue, citecolor=blue, colorlinks=true, urlcolor=blue]{hyperref}
\hypersetup{colorlinks=true}
\usepackage{stackrel}
\usepackage{pgfplots}
\usepackage{subfigure}
\usepackage{makecell}
\usetikzlibrary{positioning}
\usetikzlibrary{snakes}
\usepackage{multirow}
\usepackage{slashed}
\usepackage{float}

\newcommand\be{\begin{equation}}
\newcommand\ee{\end{equation}}
\newcommand\bea{\begin{eqnarray}}
\newcommand\eea{\end{eqnarray}}

\begin{document}

\title{ The Axion Helical Misalignment Mechanism}
\author{Wei Chao}
\email{chaowei@bnu.edu.cn}
\author{Chang-Jie Dai}
\email{daicj@bnu.edu.cn}
\affiliation{Key Laboratory of Multi-scale Spin Physics, Ministry of Education, Beijing Normal University, Beijing 100875, China}
\affiliation{Center of Advanced Quantum Studies, School of Physics and Astronomy, Beijing Normal University, Beijing, 100875, China }
\vspace{3cm}

\begin{abstract}

Understanding axion production in the early Universe remains a pivotal challenge, given the axion’s status as a compelling cold dark matter candidate. Conventional misalignment scenarios often overlook the possibility that a large initial axion velocity can fundamentally reshape the subsequent evolution of the axion field. In this Letter, we provide a comprehensive analysis of how primordial magnetic fields  impact the axion relic abundance. By accounting for the axion coupling to the Chern–Simons term of the ${\rm U}(1)_Y$ hypercharge gauge field, the axion's equation of motion is recast as a driven-oscillator equation. This modification effectively shifts the onset of axion oscillations, leading to a significant re-evaluation of the final relic abundance—a novel effect we term the axion helical misalignment mechanism. Furthermore, in the presence of primordial chiral asymmetries, the chiral magnetic effect (CME) emerges as a critical driver of axion dynamics. The interplay between the axion field and the CME not only profoundly influences the evolution of Standard Model chiral fermions but also provides a viable pathway for generating the observed baryon asymmetry of the Universe.

\end{abstract}

\maketitle
\section{Introduction}{}  \label{sec:Intro}  
 As pseudo-Nambu-Goldstone bosons associated with the spontaneous breaking of the Peccei-Quinn (PQ) symmetry~\cite{Peccei:1977hh,Peccei:1977ur,Weinberg:1977ma,Wilczek:1977pj}, axions provide an elegant dynamical solution to the strong CP problem and are among the most compelling candidates for cold dark matter (DM)~\cite{Kawasaki:2013ae,Marsh:2015xka,DiLuzio:2020wdo}. Their cosmological evolution differs significantly depending on the timing of PQ symmetry breaking relative to inflation, yielding distinct signatures in relic abundance, isocurvature fluctuations, and topological defect formation~\cite{Turner:1990uz}.
 
The misalignment mechanism is the canonical framework for generating axion DM~\cite{PRESKILL1983127,Dine1982ah,Abbott1982af,Dine:1982ah,Marsh:2015xka}. Here, the axion field is initially misaligned from the minimum of its low-energy potential. As the temperature drops and nonperturbative QCD effects generate a cosmologically relevant axion mass, the field initiates coherent oscillations. These oscillations evolve as non-relativistic matter, accounting for a portion of the observed DM density. The final relic abundance is sensitive not only to the decay constant $f_\phi$ and initial misalignment angle $\theta_i$, but also to anharmonic corrections and the temperature-dependent evolution of the axion mass~\cite{Visinelli:2009zm,Marsh:2015xka,DiLuzio:2020wdo}.
In the post-inflationary PQ breaking scenario, the relic abundance typically receives contributions not only from vacuum realignment but also from the formation, evolution, and eventual decay of axionic cosmic strings and domain walls~\cite{Ringwald:2018zjv,Gorghetto:2018myk,Gorghetto:2020qws,Buschmann:2019icd}. Recent numerical simulations have demonstrated that these topological defects can contribute substantially to the total abundance, potentially shifting the preferred QCD axion mass window~\cite{Gorghetto:2018myk,Gorghetto:2020qws,Buschmann:2019icd}. Consequently, it is crucial to identify mechanisms that can modify the onset of axion oscillations; even a modest delay in this timing can result in a significant change in the final relic abundance, thereby altering the viable parameter space for axion DM.

Recent years have witnessed significant progress in understanding the misalignment mechanism, with notable developments including the trapped misalignment~\cite{DiLuzio:2021gos,DiLuzio:2024fyt}, kinetic misalignment~\cite{Co:2019jts,Chang:2019tvx,Barman:2021oek}, and hill-top misalignment mechanisms~\cite{Co:2018mho,Co:2024bme,Takahashi:2019ehq}, among others. These studies have fundamentally reshaped the relationship between relic abundance, the decay constant, and the axion mass, effectively extending the viable parameter space for DM production beyond the constraints of the conventional misalignment framework.
In this Letter, we investigate axion production in the presence of a helical primordial magnetic field (PMF). In the early Universe, axions generically couple to the Chern–Simons term of the hypercharge gauge field—a fundamental interaction in axion electrodynamics, hypermagnetogenesis, and anomaly-induced transport phenomena~\cite{Carosi2013rla,Marsh:2015xka,Setabuddin2025vlc,Adshead:2016iae,Anber:2006xt}. When a helical PMF is present, the coherent oscillation of the axion field deviates from the standard homogeneous misalignment dynamics, as it is driven by an additional source term tied to magnetic helicity. Consequently, determining the correct relic abundance requires the simultaneous evolution of both the magnetic helicity and the magnetic energy density~\cite{Kamada:2017sxh,Joyce:1997uy,Giovannini:1997gp,Dvornikov:2021jpk,Durrer:2013pga}.
Furthermore, the early Universe may have harbored chiral fermion asymmetries, and it is observed to be manifestly matter–antimatter asymmetric~\cite{Sakharov1967dj,Morrissey2012db,Joyce:1997uy,Boyarsky:2011uy}. Consequently, the chiral magnetic effect (CME)~\cite{Vilenkin:1980fu,PhysRevD.78.074033,Boyarsky:2011uy,Akamatsu:2013pjd,Rogachevskii:2017uyc} may play a pivotal role in the coupled evolution of axions, chiral asymmetries, and helical magnetic fields. It may also facilitate the generation of the observed baryon asymmetry of the Universe (BAU) via anomaly-driven hypermagnetic dynamics~\cite{Giovannini:1997gp,Joyce:1997uy,Kamada:2016eeb,Long:2016lmj,Brandenburg:2021lnj}. Our results demonstrate that a sufficiently strong PMF can induce a large initial axion velocity, thereby delaying or altering the onset of coherent oscillations; this, in turn, modifies the viable parameter space ($f_\phi,~m_\phi$) and potentially provides a unified mechanism for generating the observed BAU~\cite{Kamada:2017sxh,Kamada:2016eeb,Long:2016lmj,Brandenburg:2021lnj}. 

Modern cosmological observations impose stringent constraints on these production mechanisms, ranging from cosmic microwave background (CMB) measurements and probes of PMF via large-scale structure and ionization history, to gravitational-wave searches and direct axion detection experiments~\cite{Durrer:2013pga,Planck:2015zrl,Paoletti:2022pfi,Lasky:2015lej,Caprini:2018mtu,Marsh:2015xka}. Furthermore, forthcoming observations from axion haloscopes like ADMX~\cite{ADMX2018gho}, dielectric and broadband searches such as MADMAX and DMRadio~\cite{MADMAX:2022urh,DMRadio:2022pkf}, helioscopes including CAST and IAXO~\cite{CAST2017uph,IAXO:2019mpb}, and next-generation gravitational-wave interferometers~\cite{LIGOScientific:2016aoc,Caprini:2019egz} will provide unprecedented opportunities to discriminate between various production scenarios. Ultimately, elucidating axion production in magnetized chiral plasmas not only addresses the DM puzzle but also offers a window into physics beyond the Standard Model (SM) by unifying axion cosmology, primordial magnetogenesis, anomaly-induced transport, and baryogenesis within a coherent early-Universe framework~\cite{Marsh:2015xka,Durrer:2013pga,Boyarsky:2011uy,Kamada:2016eeb}.


\section{The Helical Misalignment Mechanism}\label{sec:Hmm} 

We consider the dynamic of the axion $\phi$ with decay constant $f_\phi$ in the early Universe. The effective action describing the axion-photon interaction takes the following form~\cite{Setabuddin2025vlc}
\begin{align}
\begin{small}
S=\int d^4 x \sqrt{-g} \left[ \frac{1}{2} ( \partial_\mu \phi ) (\partial^\mu \phi ) -V(\phi) -\frac{g_{\phi \gamma }}{4}  \phi F_{\mu\nu }^{} \widetilde{F}^{\mu \nu}_{}\right]
\end{small}
\end{align}
where $g = \det(g_{\mu\nu})$ is the determinant of the metric tensor, $F^{\mu\nu}$ is the field strength tensor of the $\mathrm{U}(1)_{\mathrm{EM}}$ gauge field, and $g_{\phi\gamma}$ is the axion-photon coupling coefficient. Note that, prior to the electroweak phase transition, $F^{\mu\nu}$ corresponds to the field strength tensor of the $\mathrm{U}(1)_Y$ gauge field.

The equation of motion (EOM) governing the axion-like field in the Friedmann-Robertson-Walker (FRW) cosmology is~\cite{Turner1987bw,Setabuddin2025vlc} 
\begin{eqnarray}
\label{phi EoM}
\partial^2_{\eta}\phi+2\mathcal{H}\partial_{\eta}\phi+a^2m_\phi^2 \phi=g_{\phi\gamma}a^{-2} \bold{E}^*\cdot  \bold{B}^*~,
\end{eqnarray}
where $a$ is the scale factor, $\eta$ is the conformal time, $\mathcal{H} = \partial_\eta a / a$ is the conformal Hubble parameter, and $\mathbf{E}^*=a^2\mathbf{E}$ and $\mathbf{B}^*=a^2\mathbf{B}$ denote the electric and magnetic fields in the conformal frame, respectively.  In the conventional misalignment mechanism  \cite{PRESKILL1983127,Dine1982ah,Abbott1982af}, one assumes $\partial_\eta \phi_i \equiv0 $, neglects PMF, and treats the axion-like field as frozen at a spatially random initial value. 
By contrast, in the kinetic misalignment mechanism~\cite{Co2019jts,Chang2019tvx}, it has been shown that $\partial_\eta \phi_i$ can remain non-zero if the Lagrangian contains explicit breaking terms of the PQ symmetry. Consequently, the axion relic abundance may be significantly enhanced or suppressed relative to the conventional prediction.

We investigate axion production in the presence of PMFs. Magnetic fields are ubiquitous on cosmological scales, ranging from micro-gauss-strength fields coherent over tens of kiloparsecs in galaxies and galaxy clusters, to ultra-weak fields ($\sim 10^{-16}\,\mathrm{G}$) pervading intergalactic voids on megaparsec scales~\cite{Subramanian2015lua}. Their widespread existence in diffuse cosmic structures, combined with the challenge of accounting for such large-scale, volume-filling fields solely through late-time astrophysical processes, strongly motivates the hypothesis of a primordial origin~\cite{Tajima,Turner1987bw,Grasso2000wj}.
The generation of PMF is intimately tied to symmetry-breaking phenomena and quantum fluctuations during the early Universe. Two leading theoretical frameworks dominate current discussions: inflationary magnetogenesis~\cite{Turner1987bw} and phase-transition-induced production~\cite{Vachaspati1991nm,Cornwall1997ms}.

After their generation, PMF evolves dynamically through the the radiation-dominated era and subsequent cosmic epochs, shaped by interactions with the primordial plasma and cosmic expansion \cite{Subramanian2015lua,PhysRevD.57.3264,Durrer2013pga,Widrow2002ud}. In the early radiation dominated Universe, PMF is tightly coupled to the ionized plasma, driving vortical motion and magnetohydrodynamic (MHD) turbulence \cite{Brandenburg2004AstrophysicalMF,Banerjee2004df,Brandenburg2016odr}. Their linear evolution can be described by the MHD equations in an expanding background, with viscous effects damping small-scale perturbations while larger-scale magnetic structures remain comparatively robust \cite{Jedamzik1996wp,Subramanian1997gi}. As the Universe cools below the QCD scale, the plasma composition changes \cite{Schwarz2009ii}, and magnetic field evolution transitions to a regime dominated by Alfv\'{e}n waves and magnetosonic modes \cite{Kahniashvili2010wm}.  Due to the existence of the axion-photon interaction, PMF can also change the evolution of the axion in the early Universe \cite{Raffelt1987im,Mirizzi2007hr,Marsh2015xka}.

To investigate the impact of the PMF on the axion production, we explicitly take into account the evolution of the conformal helicity $h$, which is defined as
\begin{align}
h^*=\lim_{V^*\rightarrow \infty } \frac{1}{V^*} \int d^3x \bold{A}^*\cdot \bold{B}^*~,
\end{align}
where \(V^*=a^{-3}V\) denotes the  comoving volume, and \(\mathbf{A}^{*}=a\mathbf{A}\) is the vector potential of the conformal electromagnetic field \cite{Subramanian2015lua}. The corresponding evolution equation is given by
\begin{equation}
	\label{partial h}
	\frac{\partial h^*}{\partial \eta}
	=
	\lim_{V^*\to\infty}\frac{1}{V^*}\int d^3x\,
	\left(-2\,\mathbf{E}^{*}\cdot\mathbf{B}^{*}\right).
\end{equation}

To further simplify this expression, we invoke the Amp\`ere--Maxwell law in the MHD limit, where the displacement current can be neglected, i.e., \(\partial_{\eta}\mathbf{E}^{*}\approx 0\) \cite{Subramanian2015lua}. One then obtains
$\nabla\times\mathbf{B}^{*}=\mathbf{J}^{*}$,  where $\mathbf{J}^*=a^3\mathbf{J}$ denotes conformal current. The generalized Ohm's law can be written as
\begin{equation}
	\mathbf{J}^{*}
	=
	\sigma^{*}\left(\mathbf{E}^{*}+\mathbf{v}\times\mathbf{B}^{*}\right)
	-
	g_{\phi\gamma}\mathbf{B}^{*}\,\partial_{\eta}\phi
	+
	\frac{2\alpha}{\pi}\mu_{5}^{*}\mathbf{B}^{*},
\end{equation}
where \(\sigma^{*}\) denotes the conformal plasma conductivity, \(\mathbf{v}\) is the fluid velocity obeying the Navier--Stokes equation, and \(\mu_{5}^{*}\) denotes the appropriate combination of conformal chiral chemical potentials of the SM particles. The last term represents the CME \cite{PhysRevD.78.074033}, namely, the generation of an electric current along an external magnetic field in a chirally asymmetric medium.

Substituting the generalized Ohm's law into the evolution equation of \(h^*\), and using \((\mathbf{v}\times\mathbf{B}^{*})\cdot\mathbf{B}^{*}=0\), \(\mathbf{J}^{*}=\nabla\times\mathbf{B}^{*}\), \(\mathbf{B}^{*}=\nabla\times\mathbf{A}^{*}\), and adopting the Coulomb gauge \(\nabla\cdot\mathbf{A}^{*}=0\), such that \(\mathbf{J}^{*}=-\nabla^{2}\mathbf{A}^{*}\),  we obtain
\begin{align}
	\frac{\partial h^*}{\partial \eta}
	=
	\lim_{V^*\to\infty}\frac{1}{V^*}\int d^3x\,
	&\frac{2}{\sigma^{*}}
	\left(\nabla^{2}\mathbf{A}^{*}\right)\cdot\mathbf{B}^{*}
	\notag\\ +&\frac{1}{\sigma^*}
	\Big(
	\frac{8\alpha}{\pi}\mu_{5}^{*}
	-
	{4g_{\phi\gamma}}\partial_{\eta}\phi
	\Big)\rho_{B}^*,
	\label{hEoM}
\end{align}
where  $\rho_B^*$ is the conformal energy density  of the magnetic field, defined as 
\begin{equation}
\rho_B^*= \lim_{V\to \infty} \frac{1}{V^*}\int d^3 x \frac{|\mathbf{B}^*|^2}{2}  \; .
\end{equation}
Following analogous steps, we obtain the evolution equation for $\rho_B^*$
\begin{align}
\frac{\partial \rho_B^*}{\partial \eta}=\lim_{V^*\rightarrow\infty}- &\frac{1}{V^*}\frac{1}{\sigma^*} \int d^3x \bold{B}^* \cdot \Big[ \nabla^2\bold{B}^*\notag\\+&\nabla\times \Big(-\frac{2\alpha}{\pi}\mu_5^*\bold{B}^*
+g_{\phi\gamma}\bold{B}^* \partial_\eta \phi\Big) \Big]~. \label{rhoEoM}
\end{align}

\begin{figure*}[t] 
 	\centering 
 	\includegraphics[width=0.485\textwidth]{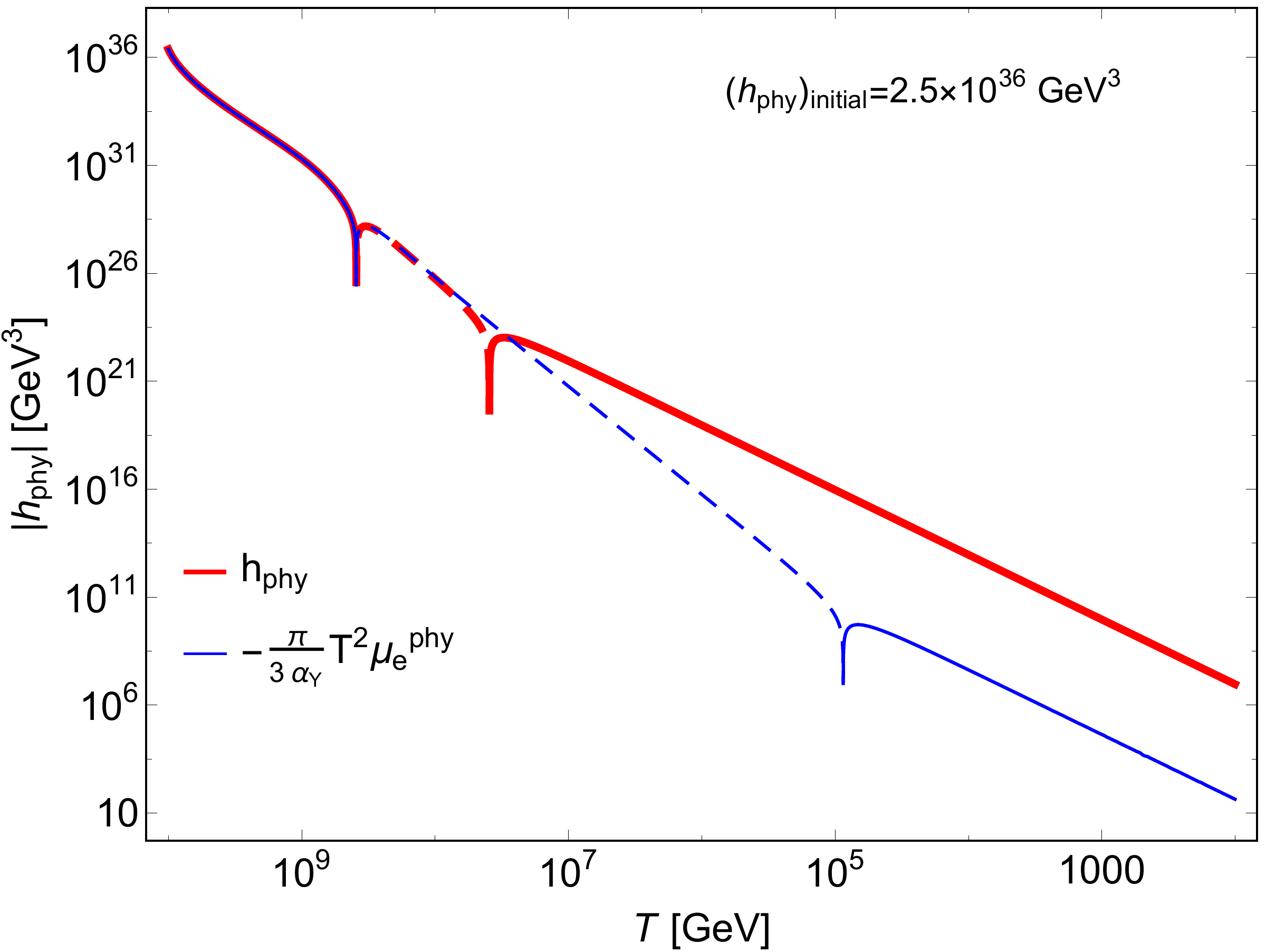} 
	\includegraphics[width=0.485\textwidth]{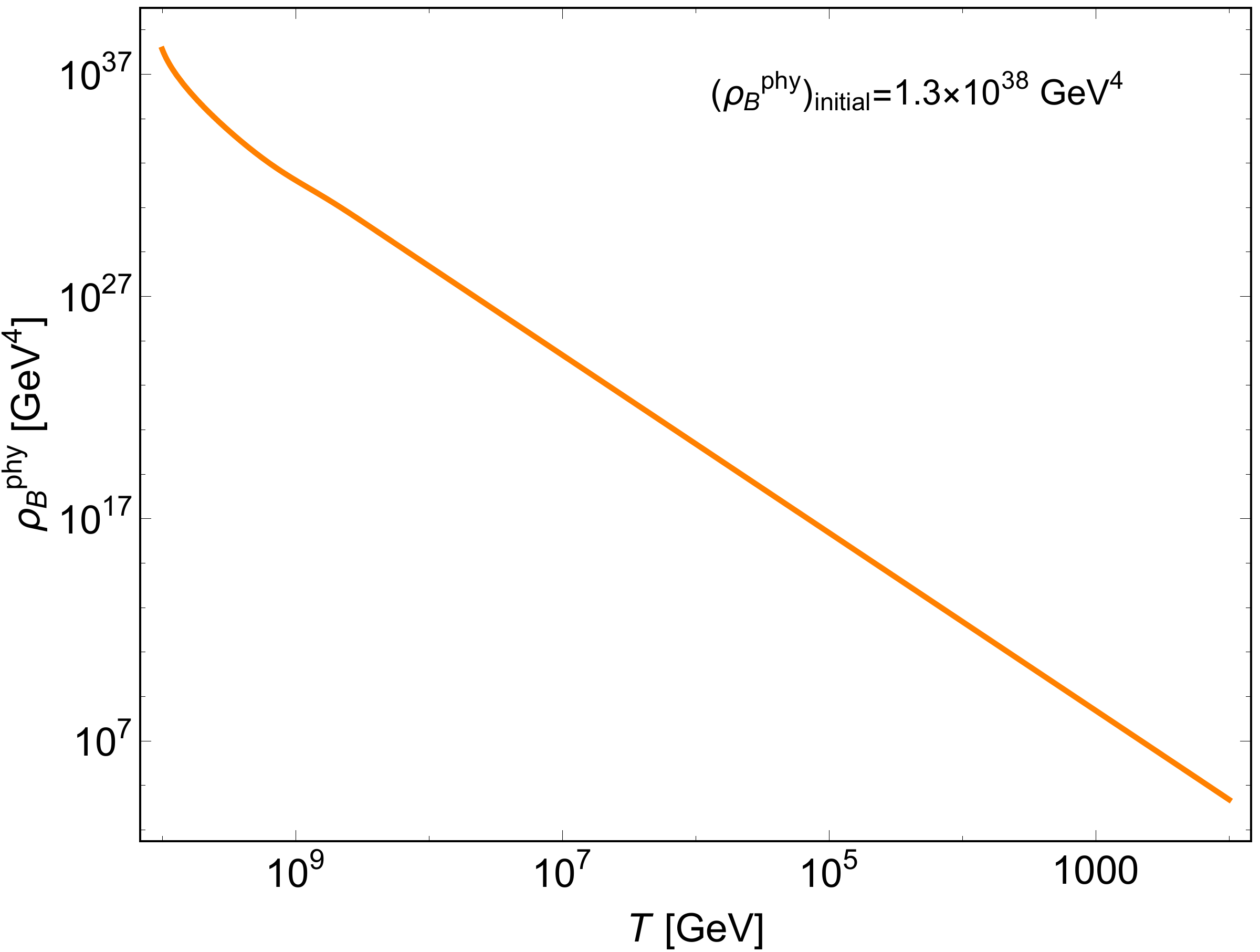} 
 	\caption{Left-panel: Evolution of total physical helicity $h_{\rm phy}$ and the physical right-handed electron chemical potential $\mu_e^{\rm phy}$. The dashed line indicates negative values; Right-panel: Evolution of total physical energy density $\rho_{B}^{\rm phy}$ as the function of the  temperature. } 
 	\label{fig:h} 
 \end{figure*}
 
We have thus derived the evolution equations for the helicity and energy density of the PMF in the presence of the CME and the axion field. Meanwhile, the EOM for the axion is given in Eq.~\ref{phi EoM}, where the source term $g_{\phi\gamma}a^{-2} \bold{E}^*\cdot  \bold{B}^*$ on the right-hand side, which governs the axion evolution, can be expressed in terms of the helicity derivative $\partial_\eta h^*$ via Eq.~\ref{partial h}. It then remains to determine the evolution of the conformal chiral chemical potential $\mu_5^*$. Assuming all interactions except that of the right-handed electron Yukawa interaction are in thermal equilibrium, one has~\cite{Domcke2019mnd}
\begin{align}
\mu_5^* =\frac{711}{481} \mu_{e}^*~,
\end{align}
where $\mu_{e}^*$ denotes  conformal chemical potential of right-handed electron and the evolution of $\mu_e^*$ is governed by \cite{PhysRevD.94.123509,PhysRevLett.79.1193,PhysRevD.94.063501}
\begin{align}
\frac{\partial \mu_e^* }{\partial \eta} = -\frac{3}{(T^*)^2 }\frac{\alpha}{\pi } \frac{\partial h}{\partial \eta } -\Gamma_{\rm Y_e} \frac{711}{481} \mu_e^*~, \label{mue}
\end{align}
where \(T^*\) is the comoving temperature, \(\Gamma_{Y_e}\approx 9.2\times10^{-14}T^*\) is the rate of the electron Yukawa interaction \cite{Bodeker_2019}, and \(\alpha=0.015\) is the fine-structure constant associated with the hypercharge gauge field \cite{Domcke2019mnd}. The first term on the right-hand side originates from the evolution of the magnetic helicity and has not been taken into account in previous studies.

In actual calculations, however, it is more convenient to perform a Fourier transformation of the electromagnetic field and consider the evolution equations for each  conformal momentum mode \(k^*\) ($k^*=a k$). Accordingly, the total  conformal energy density and helicity can be expressed as integrals over the corresponding spectral densities, $h^* =\int dk^* ~h_k^*$ and $\rho_B^* =\int dk^* ~\rho_{B,k}^*$, respectively, where the $h_k^*$ and $\rho_{B,k}^*$ are defined from the Fourier-transformed vector potential $(\bold{A}^*)_k^{\pm}$ of the electromagnetic field 
\begin{align}
h_k^*&=\frac{(k^*)^3}{2\pi^2}\left(
|(\bold{A}^*)_k^+|^2- |(\bold{A}^*)_k^-|^2
\right)~, \\
\rho_{B,k}^*&=\frac{(k^*)^4}{4\pi^2}\left(
|(\bold{A}^*)_k^+|^2+ |(\bold{A}^*)_k^-|^2
\right)~,
\end{align}
The evolution equations of $h_k^*$ and $\rho_{B,k}^*$ are given by
\begin{align}
	\label{hk}
		\frac{\partial h_k^*}{\partial \eta}&=
	-\frac{2(k^*)^2}{\sigma^*} h_k^*  
	+\frac{1}{\sigma^*}
	\left(\frac{8\alpha}{\pi}\mu_5^*-4g_{\phi\gamma} \partial_\eta \phi \right)
	\rho_{B,k}^*~,\\
	\label{rhok}
	\frac{\partial \rho_{B,k}^*}{\partial \eta}&=
	-\frac{2(k^*)^2}{\sigma^*} \rho_{B,k} ^*
	+\frac{1}{\sigma^*}
	\left(\frac{2\alpha}{\pi}\mu_5^*-g_{\phi\gamma} \partial_\eta \phi \right)
	(k^*)^2h_k^*~,
\end{align}

We proceed by numerically solving the coupled evolution equations.
%
%
We first specify the parameters adopted for our numerical analysis. Assuming a radiation-dominated Universe, the Hubble parameter is given by $H=T^2/M_*$, with  $M_*\simeq 7.11 \times 10^{17}\,\mathrm{GeV}$~\cite{1990The, DiLuzio2020wdo}. The scale factor evolves as  $T^*=aT$, where the comoving temperature is fixed at $T^* = 2 \times 10^{15}\,\mathrm{GeV}$. Consequently, the conformal time $\eta$ and the conformal Hubble parameter $\mathcal{H}$ are related by $\eta = M_*/(T^*T)$ and $\mathcal{H} = 1/\eta$. For the axion field, we adopt the initial conditions $\theta_i = \pi/\sqrt{3}$ and $(\partial_\eta \theta)_i = 0$~\cite{DiLuzio2020wdo}, and set the coupling constant and decay constant to $g_{\phi\gamma} = 10^{-12}\,\mathrm{GeV}^{-1}$ and $f_\phi = 10^{10}\,\mathrm{GeV}$, respectively.
 For the axion mass, we follow Ref.~\cite{Co2019jts} and take
 \begin{equation}
 	m_{\phi}(T)=0.6~{\rm meV}\left(\frac{\Lambda_{\rm QCD}}{T}\right)^{4},
 	\qquad T>\Lambda_{\rm QCD},
 \end{equation}
 where \(\Lambda_{\rm QCD}=160~{\rm MeV}\), while for \(T\leq \Lambda_{\rm QCD}\) we simply set \(m_{\phi}=0.6~{\rm meV}\). 
 Although this choice is motivated by the QCD axion, the formalism can be straightforwardly generalized to axion-like particles with different masses and couplings.

Following Ref.~\cite{Domcke:2019mnd}, we impose a cutoff on the comoving momentum spectrum and write the conformal helicity as
 \begin{equation}
 	h^*=\int_{10^{-4}k_{\rm UV}^*}^{k_{\rm UV}^*} h_k^*\,dk^*,
 \end{equation}
 where the ultraviolet cutoff is given by $k_{\rm UV}^*=H_{\rm rh}{T^{*}}/{T_{\rm rh}}$, with \(H_{\rm rh}\) denoting the Hubble parameter at the end of reheating, i.e., at the onset of the radiation-dominated era.  The conformal plasma conductivity is taken to be $\sigma^{*}=70\,T^{*}$, following Ref.~\cite{Domcke:2019mnd}.  To obtain sizable effects, we set the initial  helicity as $h_{\rm rh}^{\rm phy}=2.5\times10^{36}~{\rm GeV}^{3} $ and $h_{\rm rh}^* =a^3 h_{\rm rh}^{\rm phy}$, generated from the axion inflation. In this case, the chiral anomaly relates the generated Chern-Simons charge to the chiral asymmetry. Accordingly, the initial condition for \(\mu_e^*\) is taken to be \cite{Domcke:2019mnd}
\begin{equation}
	(\mu_e^*)_i
	=
	-\frac{3}{(T^*)^2}\frac{\alpha}{\pi}h_{\rm rh}^*.
	\label{axioninf}
\end{equation}
In the Fig.~\ref{fig:h}, we present the numerical evolution of the physical helicity $h_{\rm phy}$, physical energy density $\rho_B^{\rm phy}$ and physical chemical potential $\mu_e^{\rm phy}$. 
The plot in the left-panel of Fig.~\ref{fig:h} shows the evolution of the total physical helicity \(h_{\rm phy}\) together with \(-\frac{\pi}{3\alpha_Y}T^2\mu_e^{\rm phy}\). For temperatures above \(10^8~\mathrm{GeV}\), the two quantities evolve almost identically. The total physical helicity \(h_{\rm phy}\) first changes sign and becomes negative, after which it rapidly becomes positive again and subsequently decreases with decreasing temperature. A similar behavior is found for \(-\frac{\pi}{3\alpha_Y}T^2\mu_e^{\rm phy}\). 
The plot in the right-panel of Fig.~\ref{fig:h} presents the evolution of the total physical magnetic energy density \(\rho_B^{\rm phy}\), which exhibits a monotonic decrease.
The numerical evolution is assumed to start at the end of reheating with \(T_{\rm rh}=10^{10}~{\rm GeV}\). The corresponding Hubble parameter \(H_{\rm rh}\) is much smaller than \(H_{\rm inf}=10^{14}~{\rm GeV}\), where \(H_{\rm inf}\) denotes the Hubble parameter at the end of inflation. This hierarchy implies that the reheating stage cannot be treated within the instantaneous reheating approximation.

 \begin{figure*}[t] 
 	\centering 
 	\includegraphics[width=0.485\textwidth]{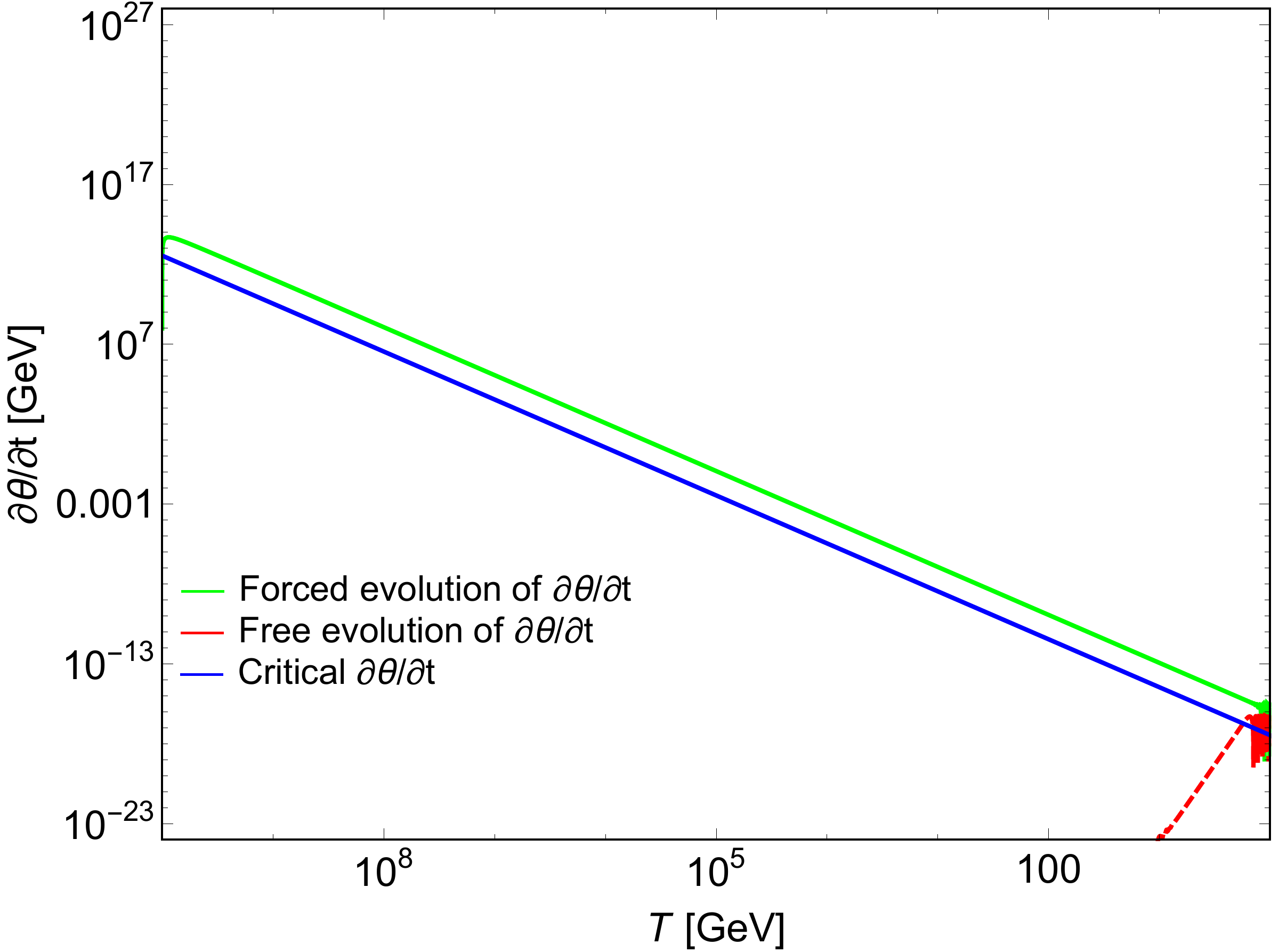} 
 	\includegraphics[width=0.485\textwidth]{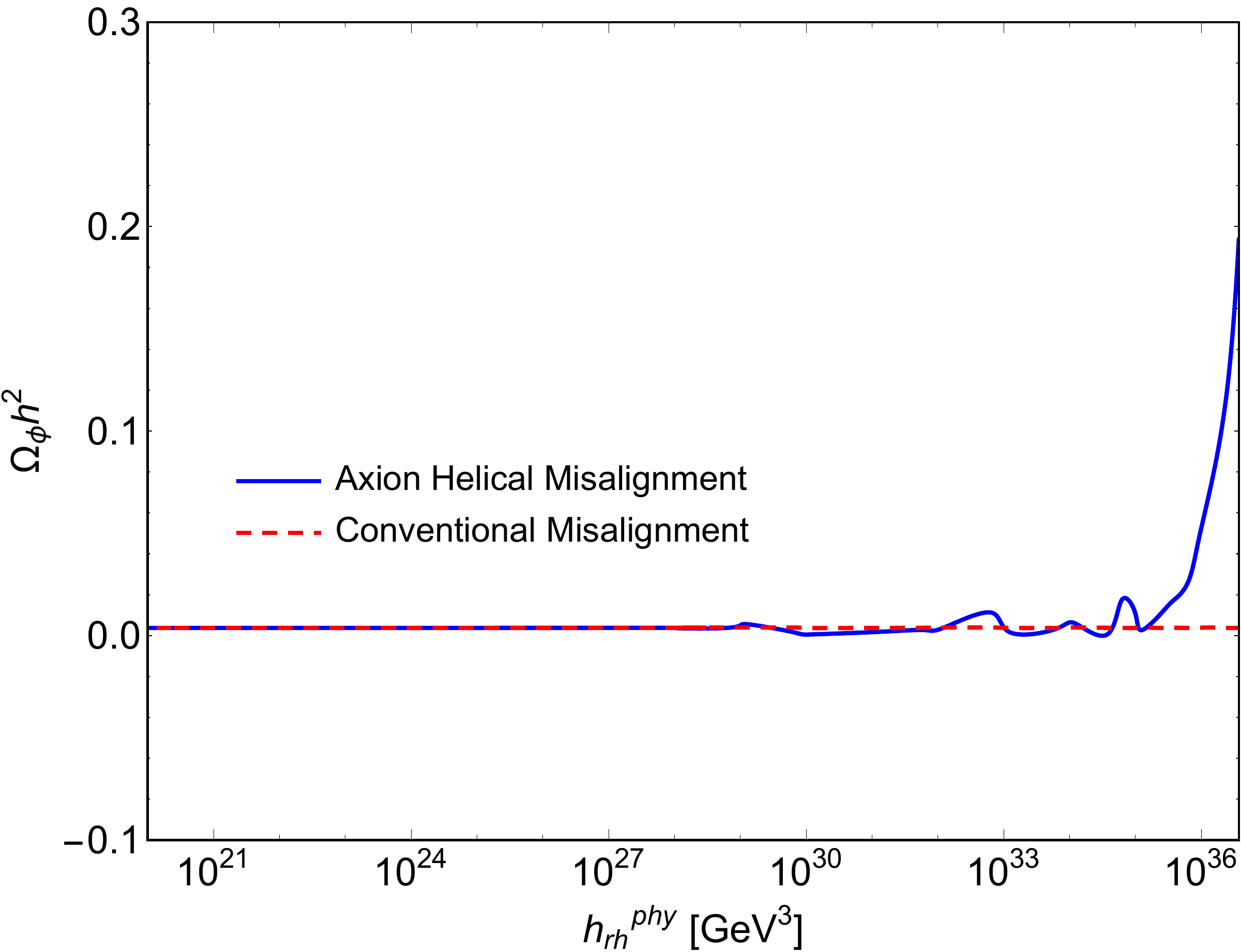} 
 	\caption{Left-panel: Evolution of $\partial_t\theta$ as the function the temperature; Right-panel: Relic abundance of the axion as the function of the initial physical helicity $h_{\rm rh}^{\rm phy}$.} 
 	\label{fig:comparsion} 
 \end{figure*}
 
The plot in the left-panel of Fig.~\ref{fig:comparsion} shows the evolution of $\partial_t\theta$ as a function of temperature (green line). The blue solid line denotes the critical velocity, defined as the minimum velocity required in the kinetic misalignment mechanism to delay the onset of oscillation \cite{Co2019jts}. One can see that the velocity obtained in our calculation consistently exceeds this critical value, thereby causing the oscillation to start later than in the  conventional evolution case (red line).

The relic abundance of the axion is given by $\Omega_\phi h^2 = h^2 {\rho_\phi}/{\rho_{\rm crit}}\,  $, where $h$ denotes the dimensionless Hubble parameter, $H_0/(100\,{\rm km\,s^{-1}\,Mpc^{-1}})$~\cite{Planck2018nkj}, rather than the magnetic helicity. The critical density is given by $\rho_{\rm crit} = {3H_0^2}/{(8\pi G)}$, and the present-day axion energy density is given by \cite{DiLuzio:2020wdo}
\begin{align}
	\rho_\phi (T_0) = \frac{T_{\rm osc} T_0^3}{\Lambda_{\rm QCD}^4 } \frac{g_{s}(T_0)}{g_{s} (T_{\rm osc})} \rho_\phi (T_{\rm osc})~ \; ,
\end{align}
where $T_0 \sim 2.35\times 10^{-4}~{\rm eV}$ is the temperature of the Universe today, and the degrees of freedom of the entropy at present are $g_{s}(T_0)=3.91$ \cite{DiLuzio:2020wdo}. Given the numerical results for the oscillation temperature $T_{\rm osc}$ and the corresponding energy density $\rho_\phi(T_{\rm osc})$, one can determine the final axion relic abundance as a function of the $h_{\rm rh}^{\rm phy}$, which has not been investigated previously.

The plot in the right-panel of  the Fig.~\ref{fig:comparsion}, shows the relic abundance of the axion as the function of the $h_{\rm rh}^{\rm phy}$.  For the parameter setting $f_\phi=10^{10}~{\rm GeV}$ and $m_\phi=0.6~{\rm meV}$, the conventional misalignment mechanism gives a negligible relic abundance as shown by the horizontal dashed line, which corresponds to take $g_{a\gamma} =0$ in solving the Eq.~\ref{phi EoM}. While the blue solid line is derived by taking $g_{a\gamma} =10^{-12}~{\rm GeV}$. It demonstrates that the axion relic abundance can be significantly modified by the PMF.  When $h_{\rm rh}^{\rm phy}<10^{28}~{\rm GeV}^3$, the abundances predicted by the conventional misalignment mechanism and the axion helical misalignment mechanism are in complete agreement. This corresponds to the regime in which the initial axion velocity $\partial_t\theta$ remains below the critical velocity in the kinetic misalignment picture, so that no delay of the oscillation onset occurs.
For larger values of $h_{\rm rh}^{\rm phy}$, the maximum velocity reached by the axion becomes close to the critical velocity, and the resulting abundance exhibits fluctuations relative to that of the conventional mechanism. As $h_{\rm rh}^{\rm phy}$ increases further, the axion velocity substantially exceeds the critical value, leading to a pronounced delay in the onset of oscillation. Consequently, the relic abundance rises rapidly with increasing $h_{\rm rh}^{\rm phy}$.
As can be read from the plot, the axion abundance reaches $\Omega_\phi h^2 \sim 0.12$ for $f_\phi=10^{10}~{\rm GeV}$, $m_\phi =0.6 ~{\rm meV}$ and $h_{\rm rh}^{\rm phy}=2.5\times10^{36}~{\rm GeV}^{3}$, which shows that the available parameter space $(f_\phi, m_\phi)$ can be extended in this mechanism.
\section{Discussion}{} 
\label{sec:summ}

The axion is a compelling cold DM candidate, and understanding its production in the early Universe remains a pivotal task. Conventional misalignment scenarios often fail to systematically account for critical factors such as the axion’s initial field value and velocity, which have been addressed in many references. In this work, we have investigated the impact of helical PMFs on the axion relic abundance. We show that, in the presence of a PMF, the axion equation of motion takes the form of a driven-oscillator equation; this modification effectively shifts the onset of axion oscillations, thereby significantly altering the final relic abundance. We term this novel effect the axion helical misalignment mechanism. Furthermore, the interplay between the axion field and the CME profoundly influences the evolution of SM chiral fermions, providing a viable pathway for the generation of the observed BAU. 
It should be noted that a systematic study of the evolution of the various transport equations for SM chiral fermions—incorporating all spectator effects and those induced by axiogenesis mechanism—is beyond the scope of this work. Furthermore, the potential impact of initial chemical potentials on the final axion relic abundance is not addressed here. We will present a comprehensive analysis of these effects in a forthcoming paper

\begin{acknowledgments}

This work was supported in part by the National Key R\&D Program of China under Grant No. 2023YFA1607104, by the National Natural Science Foundation of China under Grants No. 12447105.

\end{acknowledgments}

\bibliography{references}

\begin{thebibliography}{85}%
\makeatletter
\providecommand \@ifxundefined [1]{%
 \@ifx{#1\undefined}
}%
\providecommand \@ifnum [1]{%
 \ifnum #1\expandafter \@firstoftwo
 \else \expandafter \@secondoftwo
 \fi
}%
\providecommand \@ifx [1]{%
 \ifx #1\expandafter \@firstoftwo
 \else \expandafter \@secondoftwo
 \fi
}%
\providecommand \natexlab [1]{#1}%
\providecommand \enquote  [1]{``#1''}%
\providecommand \bibnamefont  [1]{#1}%
\providecommand \bibfnamefont [1]{#1}%
\providecommand \citenamefont [1]{#1}%
\providecommand \href@noop [0]{\@secondoftwo}%
\providecommand \href [0]{\begingroup \@sanitize@url \@href}%
\providecommand \@href[1]{\@@startlink{#1}\@@href}%
\providecommand \@@href[1]{\endgroup#1\@@endlink}%
\providecommand \@sanitize@url [0]{\catcode `\\12\catcode `\$12\catcode
  `\&12\catcode `\#12\catcode `\^12\catcode `\_12\catcode `\%12\relax}%
\providecommand \@@startlink[1]{}%
\providecommand \@@endlink[0]{}%
\providecommand \url  [0]{\begingroup\@sanitize@url \@url }%
\providecommand \@url [1]{\endgroup\@href {#1}{\urlprefix }}%
\providecommand \urlprefix  [0]{URL }%
\providecommand \Eprint [0]{\href }%
\providecommand \doibase [0]{http://dx.doi.org/}%
\providecommand \selectlanguage [0]{\@gobble}%
\providecommand \bibinfo  [0]{\@secondoftwo}%
\providecommand \bibfield  [0]{\@secondoftwo}%
\providecommand \translation [1]{[#1]}%
\providecommand \BibitemOpen [0]{}%
\providecommand \bibitemStop [0]{}%
\providecommand \bibitemNoStop [0]{.\EOS\space}%
\providecommand \EOS [0]{\spacefactor3000\relax}%
\providecommand \BibitemShut  [1]{\csname bibitem#1\endcsname}%
\let\auto@bib@innerbib\@empty
\bibitem [{\citenamefont {Peccei}\ and\ \citenamefont
  {Quinn}(1977{\natexlab{a}})}]{Peccei:1977hh}%
  \BibitemOpen
  \bibfield  {author} {\bibinfo {author} {\bibfnamefont {R.~D.}\ \bibnamefont
  {Peccei}}\ and\ \bibinfo {author} {\bibfnamefont {Helen~R.}\ \bibnamefont
  {Quinn}},\ }\bibfield  {title} {\enquote {\bibinfo {title} {{CP Conservation
  in the Presence of Instantons}},}\ }\href {\doibase
  10.1103/PhysRevLett.38.1440} {\bibfield  {journal} {\bibinfo  {journal}
  {Phys. Rev. Lett.}\ }\textbf {\bibinfo {volume} {38}},\ \bibinfo {pages}
  {1440--1443} (\bibinfo {year} {1977}{\natexlab{a}})}\BibitemShut {NoStop}%
\bibitem [{\citenamefont {Peccei}\ and\ \citenamefont
  {Quinn}(1977{\natexlab{b}})}]{Peccei:1977ur}%
  \BibitemOpen
  \bibfield  {author} {\bibinfo {author} {\bibfnamefont {R.~D.}\ \bibnamefont
  {Peccei}}\ and\ \bibinfo {author} {\bibfnamefont {Helen~R.}\ \bibnamefont
  {Quinn}},\ }\bibfield  {title} {\enquote {\bibinfo {title} {{Constraints
  Imposed by CP Conservation in the Presence of Instantons}},}\ }\href
  {\doibase 10.1103/PhysRevD.16.1791} {\bibfield  {journal} {\bibinfo
  {journal} {Phys. Rev. D}\ }\textbf {\bibinfo {volume} {16}},\ \bibinfo
  {pages} {1791--1797} (\bibinfo {year} {1977}{\natexlab{b}})}\BibitemShut
  {NoStop}%
\bibitem [{\citenamefont {Weinberg}(1978)}]{Weinberg:1977ma}%
  \BibitemOpen
  \bibfield  {author} {\bibinfo {author} {\bibfnamefont {Steven}\ \bibnamefont
  {Weinberg}},\ }\bibfield  {title} {\enquote {\bibinfo {title} {{A New Light
  Boson?}}}\ }\href {\doibase 10.1103/PhysRevLett.40.223} {\bibfield  {journal}
  {\bibinfo  {journal} {Phys. Rev. Lett.}\ }\textbf {\bibinfo {volume} {40}},\
  \bibinfo {pages} {223--226} (\bibinfo {year} {1978})}\BibitemShut {NoStop}%
\bibitem [{\citenamefont {Wilczek}(1978)}]{Wilczek:1977pj}%
  \BibitemOpen
  \bibfield  {author} {\bibinfo {author} {\bibfnamefont {Frank}\ \bibnamefont
  {Wilczek}},\ }\bibfield  {title} {\enquote {\bibinfo {title} {{Problem of
  Strong $P$ and $T$ Invariance in the Presence of Instantons}},}\ }\href
  {\doibase 10.1103/PhysRevLett.40.279} {\bibfield  {journal} {\bibinfo
  {journal} {Phys. Rev. Lett.}\ }\textbf {\bibinfo {volume} {40}},\ \bibinfo
  {pages} {279--282} (\bibinfo {year} {1978})}\BibitemShut {NoStop}%
\bibitem [{\citenamefont {Kawasaki}\ and\ \citenamefont
  {Nakayama}(2013)}]{Kawasaki:2013ae}%
  \BibitemOpen
  \bibfield  {author} {\bibinfo {author} {\bibfnamefont {Masahiro}\
  \bibnamefont {Kawasaki}}\ and\ \bibinfo {author} {\bibfnamefont {Kazunori}\
  \bibnamefont {Nakayama}},\ }\bibfield  {title} {\enquote {\bibinfo {title}
  {{Axions: Theory and Cosmological Role}},}\ }\href@noop {} {\bibfield
  {journal} {\bibinfo  {journal} {Ann. Rev. Nucl. Part. Sci.}\ }\textbf
  {\bibinfo {volume} {63}},\ \bibinfo {pages} {69--95} (\bibinfo {year}
  {2013})},\ \Eprint {http://arxiv.org/abs/1301.1123} {arXiv:1301.1123
  [hep-ph]} \BibitemShut {NoStop}%
\bibitem [{\citenamefont {Marsh}(2016{\natexlab{a}})}]{Marsh:2015xka}%
  \BibitemOpen
  \bibfield  {author} {\bibinfo {author} {\bibfnamefont {David J.~E.}\
  \bibnamefont {Marsh}},\ }\bibfield  {title} {\enquote {\bibinfo {title}
  {{Axion Cosmology}},}\ }\href {\doibase 10.1016/j.physrep.2016.06.005}
  {\bibfield  {journal} {\bibinfo  {journal} {Phys. Rept.}\ }\textbf {\bibinfo
  {volume} {643}},\ \bibinfo {pages} {1--79} (\bibinfo {year}
  {2016}{\natexlab{a}})},\ \Eprint {http://arxiv.org/abs/1510.07633}
  {arXiv:1510.07633 [astro-ph.CO]} \BibitemShut {NoStop}%
\bibitem [{\citenamefont {Di~Luzio}\ \emph
  {et~al.}(2020{\natexlab{a}})\citenamefont {Di~Luzio}, \citenamefont
  {Giannotti}, \citenamefont {Nardi},\ and\ \citenamefont
  {Visinelli}}]{DiLuzio:2020wdo}%
  \BibitemOpen
  \bibfield  {author} {\bibinfo {author} {\bibfnamefont {Luca}\ \bibnamefont
  {Di~Luzio}}, \bibinfo {author} {\bibfnamefont {Maurizio}\ \bibnamefont
  {Giannotti}}, \bibinfo {author} {\bibfnamefont {Enrico}\ \bibnamefont
  {Nardi}}, \ and\ \bibinfo {author} {\bibfnamefont {Luca}\ \bibnamefont
  {Visinelli}},\ }\bibfield  {title} {\enquote {\bibinfo {title} {{The
  landscape of QCD axion models}},}\ }\href {\doibase
  10.1016/j.physrep.2020.06.002} {\bibfield  {journal} {\bibinfo  {journal}
  {Phys. Rept.}\ }\textbf {\bibinfo {volume} {870}},\ \bibinfo {pages} {1--117}
  (\bibinfo {year} {2020}{\natexlab{a}})},\ \Eprint
  {http://arxiv.org/abs/2003.01100} {arXiv:2003.01100 [hep-ph]} \BibitemShut
  {NoStop}%
\bibitem [{\citenamefont {Turner}\ and\ \citenamefont
  {Wilczek}(1991)}]{Turner:1990uz}%
  \BibitemOpen
  \bibfield  {author} {\bibinfo {author} {\bibfnamefont {Michael~S.}\
  \bibnamefont {Turner}}\ and\ \bibinfo {author} {\bibfnamefont {Frank}\
  \bibnamefont {Wilczek}},\ }\bibfield  {title} {\enquote {\bibinfo {title}
  {{Inflationary axion cosmology}},}\ }\href@noop {} {\bibfield  {journal}
  {\bibinfo  {journal} {Phys. Rev. Lett.}\ }\textbf {\bibinfo {volume} {66}},\
  \bibinfo {pages} {5--8} (\bibinfo {year} {1991})}\BibitemShut {NoStop}%
\bibitem [{\citenamefont {Preskill}\ \emph {et~al.}(1983)\citenamefont
  {Preskill}, \citenamefont {Wise},\ and\ \citenamefont
  {Wilczek}}]{PRESKILL1983127}%
  \BibitemOpen
  \bibfield  {author} {\bibinfo {author} {\bibfnamefont {John}\ \bibnamefont
  {Preskill}}, \bibinfo {author} {\bibfnamefont {Mark~B.}\ \bibnamefont
  {Wise}}, \ and\ \bibinfo {author} {\bibfnamefont {Frank}\ \bibnamefont
  {Wilczek}},\ }\bibfield  {title} {\enquote {\bibinfo {title} {Cosmology of
  the invisible axion},}\ }\href {\doibase
  https://doi.org/10.1016/0370-2693(83)90637-8} {\bibfield  {journal} {\bibinfo
   {journal} {Physics Letters B}\ }\textbf {\bibinfo {volume} {120}},\ \bibinfo
  {pages} {127--132} (\bibinfo {year} {1983})}\BibitemShut {NoStop}%
\bibitem [{\citenamefont {Dine}\ and\ \citenamefont
  {Fischler}(1983{\natexlab{a}})}]{Dine1982ah}%
  \BibitemOpen
  \bibfield  {author} {\bibinfo {author} {\bibfnamefont {Michael}\ \bibnamefont
  {Dine}}\ and\ \bibinfo {author} {\bibfnamefont {Willy}\ \bibnamefont
  {Fischler}},\ }\bibfield  {title} {\enquote {\bibinfo {title} {{The Not So
  Harmless Axion}},}\ }\href {\doibase 10.1016/0370-2693(83)90639-1} {\bibfield
   {journal} {\bibinfo  {journal} {Phys. Lett. B}\ }\textbf {\bibinfo {volume}
  {120}},\ \bibinfo {pages} {137--141} (\bibinfo {year}
  {1983}{\natexlab{a}})}\BibitemShut {NoStop}%
\bibitem [{\citenamefont {Abbott}\ and\ \citenamefont
  {Sikivie}(1983)}]{Abbott1982af}%
  \BibitemOpen
  \bibfield  {author} {\bibinfo {author} {\bibfnamefont {L.~F.}\ \bibnamefont
  {Abbott}}\ and\ \bibinfo {author} {\bibfnamefont {P.}~\bibnamefont
  {Sikivie}},\ }\bibfield  {title} {\enquote {\bibinfo {title} {{A Cosmological
  Bound on the Invisible Axion}},}\ }\href {\doibase
  10.1016/0370-2693(83)90638-X} {\bibfield  {journal} {\bibinfo  {journal}
  {Phys. Lett. B}\ }\textbf {\bibinfo {volume} {120}},\ \bibinfo {pages}
  {133--136} (\bibinfo {year} {1983})}\BibitemShut {NoStop}%
\bibitem [{\citenamefont {Dine}\ and\ \citenamefont
  {Fischler}(1983{\natexlab{b}})}]{Dine:1982ah}%
  \BibitemOpen
  \bibfield  {author} {\bibinfo {author} {\bibfnamefont {Michael}\ \bibnamefont
  {Dine}}\ and\ \bibinfo {author} {\bibfnamefont {Willy}\ \bibnamefont
  {Fischler}},\ }\bibfield  {title} {\enquote {\bibinfo {title} {The
  not-so-harmless axion},}\ }\href {\doibase 10.1016/0370-2693(83)90639-1}
  {\bibfield  {journal} {\bibinfo  {journal} {Phys. Lett. B}\ }\textbf
  {\bibinfo {volume} {120}},\ \bibinfo {pages} {137--141} (\bibinfo {year}
  {1983}{\natexlab{b}})}\BibitemShut {NoStop}%
\bibitem [{\citenamefont {Visinelli}\ and\ \citenamefont
  {Gondolo}(2009)}]{Visinelli:2009zm}%
  \BibitemOpen
  \bibfield  {author} {\bibinfo {author} {\bibfnamefont {Luca}\ \bibnamefont
  {Visinelli}}\ and\ \bibinfo {author} {\bibfnamefont {Paolo}\ \bibnamefont
  {Gondolo}},\ }\bibfield  {title} {\enquote {\bibinfo {title} {{Dark Matter
  Axions Revisited}},}\ }\href@noop {} {\bibfield  {journal} {\bibinfo
  {journal} {Phys. Rev. D}\ }\textbf {\bibinfo {volume} {80}},\ \bibinfo
  {pages} {035024} (\bibinfo {year} {2009})},\ \Eprint
  {http://arxiv.org/abs/0910.3941} {arXiv:0910.3941 [astro-ph.CO]} \BibitemShut
  {NoStop}%
\bibitem [{\citenamefont {Ringwald}(2020)}]{Ringwald:2018zjv}%
  \BibitemOpen
  \bibfield  {author} {\bibinfo {author} {\bibfnamefont {Andreas}\ \bibnamefont
  {Ringwald}},\ }\bibfield  {title} {\enquote {\bibinfo {title} {{Axion mass in
  the case of post-inflationary Peccei-Quinn symmetry breaking}},}\ }\href@noop
  {} {\bibfield  {journal} {\bibinfo  {journal} {J. Phys. Conf. Ser.}\ }\textbf
  {\bibinfo {volume} {1342}},\ \bibinfo {pages} {012004} (\bibinfo {year}
  {2020})},\ \Eprint {http://arxiv.org/abs/1805.09618} {arXiv:1805.09618
  [hep-ph]} \BibitemShut {NoStop}%
\bibitem [{\citenamefont {Gorghetto}\ \emph {et~al.}(2018)\citenamefont
  {Gorghetto}, \citenamefont {Hardy},\ and\ \citenamefont
  {Villadoro}}]{Gorghetto:2018myk}%
  \BibitemOpen
  \bibfield  {author} {\bibinfo {author} {\bibfnamefont {Marco}\ \bibnamefont
  {Gorghetto}}, \bibinfo {author} {\bibfnamefont {Edward}\ \bibnamefont
  {Hardy}}, \ and\ \bibinfo {author} {\bibfnamefont {Giovanni}\ \bibnamefont
  {Villadoro}},\ }\bibfield  {title} {\enquote {\bibinfo {title} {{Axions from
  Strings: the Attractive Solution}},}\ }\href@noop {} {\bibfield  {journal}
  {\bibinfo  {journal} {JHEP}\ }\textbf {\bibinfo {volume} {07}},\ \bibinfo
  {pages} {151} (\bibinfo {year} {2018})},\ \Eprint
  {http://arxiv.org/abs/1806.04677} {arXiv:1806.04677 [hep-ph]} \BibitemShut
  {NoStop}%
\bibitem [{\citenamefont {Gorghetto}\ \emph {et~al.}(2021)\citenamefont
  {Gorghetto}, \citenamefont {Hardy},\ and\ \citenamefont
  {Villadoro}}]{Gorghetto:2020qws}%
  \BibitemOpen
  \bibfield  {author} {\bibinfo {author} {\bibfnamefont {Marco}\ \bibnamefont
  {Gorghetto}}, \bibinfo {author} {\bibfnamefont {Edward}\ \bibnamefont
  {Hardy}}, \ and\ \bibinfo {author} {\bibfnamefont {Giovanni}\ \bibnamefont
  {Villadoro}},\ }\bibfield  {title} {\enquote {\bibinfo {title} {{More Axions
  from Strings}},}\ }\href@noop {} {\bibfield  {journal} {\bibinfo  {journal}
  {SciPost Phys.}\ }\textbf {\bibinfo {volume} {10}},\ \bibinfo {pages} {050}
  (\bibinfo {year} {2021})},\ \Eprint {http://arxiv.org/abs/2007.04990}
  {arXiv:2007.04990 [hep-ph]} \BibitemShut {NoStop}%
\bibitem [{\citenamefont {Buschmann}\ \emph {et~al.}(2020)\citenamefont
  {Buschmann}, \citenamefont {Foster},\ and\ \citenamefont
  {Safdi}}]{Buschmann:2019icd}%
  \BibitemOpen
  \bibfield  {author} {\bibinfo {author} {\bibfnamefont {Malte}\ \bibnamefont
  {Buschmann}}, \bibinfo {author} {\bibfnamefont {Joshua~W.}\ \bibnamefont
  {Foster}}, \ and\ \bibinfo {author} {\bibfnamefont {Benjamin~R.}\
  \bibnamefont {Safdi}},\ }\bibfield  {title} {\enquote {\bibinfo {title}
  {{Early-Universe Simulations of the Cosmological Axion}},}\ }\href@noop {}
  {\bibfield  {journal} {\bibinfo  {journal} {Phys. Rev. Lett.}\ }\textbf
  {\bibinfo {volume} {124}},\ \bibinfo {pages} {161103} (\bibinfo {year}
  {2020})},\ \Eprint {http://arxiv.org/abs/1906.00967} {arXiv:1906.00967
  [astro-ph.CO]} \BibitemShut {NoStop}%
\bibitem [{\citenamefont {Di~Luzio}\ \emph {et~al.}(2021)\citenamefont
  {Di~Luzio}, \citenamefont {Gavela}, \citenamefont {Quilez},\ and\
  \citenamefont {Ringwald}}]{DiLuzio:2021gos}%
  \BibitemOpen
  \bibfield  {author} {\bibinfo {author} {\bibfnamefont {Luca}\ \bibnamefont
  {Di~Luzio}}, \bibinfo {author} {\bibfnamefont {Belen}\ \bibnamefont
  {Gavela}}, \bibinfo {author} {\bibfnamefont {Pablo}\ \bibnamefont {Quilez}},
  \ and\ \bibinfo {author} {\bibfnamefont {Andreas}\ \bibnamefont {Ringwald}},\
  }\bibfield  {title} {\enquote {\bibinfo {title} {{Dark matter from an even
  lighter QCD axion: trapped misalignment}},}\ }\href {\doibase
  10.1088/1475-7516/2021/10/001} {\bibfield  {journal} {\bibinfo  {journal}
  {JCAP}\ }\textbf {\bibinfo {volume} {10}},\ \bibinfo {pages} {001} (\bibinfo
  {year} {2021})},\ \Eprint {http://arxiv.org/abs/2102.01082} {arXiv:2102.01082
  [hep-ph]} \BibitemShut {NoStop}%
\bibitem [{\citenamefont {Di~Luzio}\ and\ \citenamefont
  {S{\o}rensen}(2024)}]{DiLuzio:2024fyt}%
  \BibitemOpen
  \bibfield  {author} {\bibinfo {author} {\bibfnamefont {Luca}\ \bibnamefont
  {Di~Luzio}}\ and\ \bibinfo {author} {\bibfnamefont {Philip}\ \bibnamefont
  {S{\o}rensen}},\ }\bibfield  {title} {\enquote {\bibinfo {title} {{Axion
  production via trapped misalignment from Peccei-Quinn symmetry breaking}},}\
  }\href {\doibase 10.1007/JHEP10(2024)239} {\bibfield  {journal} {\bibinfo
  {journal} {JHEP}\ }\textbf {\bibinfo {volume} {10}},\ \bibinfo {pages} {239}
  (\bibinfo {year} {2024})},\ \Eprint {http://arxiv.org/abs/2408.04623}
  {arXiv:2408.04623 [hep-ph]} \BibitemShut {NoStop}%
\bibitem [{\citenamefont {Co}\ \emph {et~al.}(2020{\natexlab{a}})\citenamefont
  {Co}, \citenamefont {Hall},\ and\ \citenamefont {Harigaya}}]{Co:2019jts}%
  \BibitemOpen
  \bibfield  {author} {\bibinfo {author} {\bibfnamefont {Raymond~T.}\
  \bibnamefont {Co}}, \bibinfo {author} {\bibfnamefont {Lawrence~J.}\
  \bibnamefont {Hall}}, \ and\ \bibinfo {author} {\bibfnamefont {Keisuke}\
  \bibnamefont {Harigaya}},\ }\bibfield  {title} {\enquote {\bibinfo {title}
  {{Axion Kinetic Misalignment Mechanism}},}\ }\href {\doibase
  10.1103/PhysRevLett.124.251802} {\bibfield  {journal} {\bibinfo  {journal}
  {Phys. Rev. Lett.}\ }\textbf {\bibinfo {volume} {124}},\ \bibinfo {pages}
  {251802} (\bibinfo {year} {2020}{\natexlab{a}})},\ \Eprint
  {http://arxiv.org/abs/1910.14152} {arXiv:1910.14152 [hep-ph]} \BibitemShut
  {NoStop}%
\bibitem [{\citenamefont {Chang}\ and\ \citenamefont
  {Cui}(2020{\natexlab{a}})}]{Chang:2019tvx}%
  \BibitemOpen
  \bibfield  {author} {\bibinfo {author} {\bibfnamefont {Chia-Feng}\
  \bibnamefont {Chang}}\ and\ \bibinfo {author} {\bibfnamefont {Yanou}\
  \bibnamefont {Cui}},\ }\bibfield  {title} {\enquote {\bibinfo {title} {{New
  Perspectives on Axion Misalignment Mechanism}},}\ }\href {\doibase
  10.1103/PhysRevD.102.015003} {\bibfield  {journal} {\bibinfo  {journal}
  {Phys. Rev. D}\ }\textbf {\bibinfo {volume} {102}},\ \bibinfo {pages}
  {015003} (\bibinfo {year} {2020}{\natexlab{a}})},\ \Eprint
  {http://arxiv.org/abs/1911.11885} {arXiv:1911.11885 [hep-ph]} \BibitemShut
  {NoStop}%
\bibitem [{\citenamefont {Barman}\ \emph {et~al.}(2022)\citenamefont {Barman},
  \citenamefont {Bernal}, \citenamefont {Ramberg},\ and\ \citenamefont
  {Visinelli}}]{Barman:2021oek}%
  \BibitemOpen
  \bibfield  {author} {\bibinfo {author} {\bibfnamefont {Basabendu}\
  \bibnamefont {Barman}}, \bibinfo {author} {\bibfnamefont {Nicol\'as}\
  \bibnamefont {Bernal}}, \bibinfo {author} {\bibfnamefont {Nicklas}\
  \bibnamefont {Ramberg}}, \ and\ \bibinfo {author} {\bibfnamefont {Luca}\
  \bibnamefont {Visinelli}},\ }\bibfield  {title} {\enquote {\bibinfo {title}
  {Qcd axion kinetic misalignment without prejudice},}\ }\href {\doibase
  10.3390/universe8120662} {\bibfield  {journal} {\bibinfo  {journal}
  {Universe}\ }\textbf {\bibinfo {volume} {8}},\ \bibinfo {pages} {662}
  (\bibinfo {year} {2022})},\ \Eprint {http://arxiv.org/abs/2111.03677}
  {arXiv:2111.03677 [hep-ph]} \BibitemShut {NoStop}%
\bibitem [{\citenamefont {Co}\ \emph {et~al.}(2019)\citenamefont {Co},
  \citenamefont {Gonzalez},\ and\ \citenamefont {Harigaya}}]{Co:2018mho}%
  \BibitemOpen
  \bibfield  {author} {\bibinfo {author} {\bibfnamefont {Raymond~T.}\
  \bibnamefont {Co}}, \bibinfo {author} {\bibfnamefont {Eric}\ \bibnamefont
  {Gonzalez}}, \ and\ \bibinfo {author} {\bibfnamefont {Keisuke}\ \bibnamefont
  {Harigaya}},\ }\bibfield  {title} {\enquote {\bibinfo {title} {{Axion
  Misalignment Driven to the Hilltop}},}\ }\href {\doibase
  10.1007/JHEP05(2019)163} {\bibfield  {journal} {\bibinfo  {journal} {JHEP}\
  }\textbf {\bibinfo {volume} {05}},\ \bibinfo {pages} {163} (\bibinfo {year}
  {2019})},\ \Eprint {http://arxiv.org/abs/1812.11192} {arXiv:1812.11192
  [hep-ph]} \BibitemShut {NoStop}%
\bibitem [{\citenamefont {Co}\ \emph {et~al.}(2024)\citenamefont {Co},
  \citenamefont {Gherghetta}, \citenamefont {Liu},\ and\ \citenamefont
  {Lyu}}]{Co:2024bme}%
  \BibitemOpen
  \bibfield  {author} {\bibinfo {author} {\bibfnamefont {Raymond~T.}\
  \bibnamefont {Co}}, \bibinfo {author} {\bibfnamefont {Tony}\ \bibnamefont
  {Gherghetta}}, \bibinfo {author} {\bibfnamefont {Zhen}\ \bibnamefont {Liu}},
  \ and\ \bibinfo {author} {\bibfnamefont {Kun-Feng}\ \bibnamefont {Lyu}},\
  }\bibfield  {title} {\enquote {\bibinfo {title} {{A light QCD axion with
  hilltop misalignment}},}\ }\href {\doibase 10.1007/JHEP09(2024)145}
  {\bibfield  {journal} {\bibinfo  {journal} {JHEP}\ }\textbf {\bibinfo
  {volume} {09}},\ \bibinfo {pages} {145} (\bibinfo {year} {2024})},\ \Eprint
  {http://arxiv.org/abs/2407.12930} {arXiv:2407.12930 [hep-ph]} \BibitemShut
  {NoStop}%
\bibitem [{\citenamefont {Takahashi}\ and\ \citenamefont
  {Yin}(2019)}]{Takahashi:2019ehq}%
  \BibitemOpen
  \bibfield  {author} {\bibinfo {author} {\bibfnamefont {Fuminobu}\
  \bibnamefont {Takahashi}}\ and\ \bibinfo {author} {\bibfnamefont {Wen}\
  \bibnamefont {Yin}},\ }\bibfield  {title} {\enquote {\bibinfo {title} {Qcd
  axion on hilltop by a phase shift of $\pi$},}\ }\href {\doibase
  10.1007/JHEP10(2019)120} {\bibfield  {journal} {\bibinfo  {journal} {JHEP}\
  }\textbf {\bibinfo {volume} {10}},\ \bibinfo {pages} {120} (\bibinfo {year}
  {2019})},\ \Eprint {http://arxiv.org/abs/1908.06071} {arXiv:1908.06071
  [hep-ph]} \BibitemShut {NoStop}%
\bibitem [{\citenamefont {Carosi}\ \emph {et~al.}(2013)\citenamefont {Carosi},
  \citenamefont {Friedland}, \citenamefont {Giannotti}, \citenamefont
  {Pivovaroff}, \citenamefont {Ruz},\ and\ \citenamefont
  {Vogel}}]{Carosi2013rla}%
  \BibitemOpen
  \bibfield  {author} {\bibinfo {author} {\bibfnamefont {G.}~\bibnamefont
  {Carosi}}, \bibinfo {author} {\bibfnamefont {A.}~\bibnamefont {Friedland}},
  \bibinfo {author} {\bibfnamefont {M.}~\bibnamefont {Giannotti}}, \bibinfo
  {author} {\bibfnamefont {M.~J.}\ \bibnamefont {Pivovaroff}}, \bibinfo
  {author} {\bibfnamefont {J.}~\bibnamefont {Ruz}}, \ and\ \bibinfo {author}
  {\bibfnamefont {J.~K.}\ \bibnamefont {Vogel}},\ }\bibfield  {title} {\enquote
  {\bibinfo {title} {{Probing the axion-photon coupling: phenomenological and
  experimental perspectives. A snowmass white paper}},}\ }in\ \href@noop {}
  {\emph {\bibinfo {booktitle} {{Snowmass 2013}: {Snowmass on the
  Mississippi}}}}\ (\bibinfo {year} {2013})\ \Eprint
  {http://arxiv.org/abs/1309.7035} {arXiv:1309.7035 [hep-ph]} \BibitemShut
  {NoStop}%
\bibitem [{\citenamefont {Setabuddin}\ \emph {et~al.}(2025)\citenamefont
  {Setabuddin}, \citenamefont {Haque}, \citenamefont {Karmakar},\ and\
  \citenamefont {Pal}}]{Setabuddin2025vlc}%
  \BibitemOpen
  \bibfield  {author} {\bibinfo {author} {\bibnamefont {Setabuddin}}, \bibinfo
  {author} {\bibfnamefont {Md~Riajul}\ \bibnamefont {Haque}}, \bibinfo {author}
  {\bibfnamefont {Rajesh}\ \bibnamefont {Karmakar}}, \ and\ \bibinfo {author}
  {\bibfnamefont {Supratik}\ \bibnamefont {Pal}},\ }\bibfield  {title}
  {\enquote {\bibinfo {title} {{Axion-Photon Conversion in FLRW with Primordial
  Magnetic Fields: Explaining the Radio Excess}},}\ }\href@noop {} {\
  (\bibinfo {year} {2025})},\ \Eprint {http://arxiv.org/abs/2509.09472}
  {arXiv:2509.09472 [astro-ph.CO]} \BibitemShut {NoStop}%
\bibitem [{\citenamefont {Adshead}\ \emph {et~al.}(2016)\citenamefont
  {Adshead}, \citenamefont {Giblin}, \citenamefont {Scully},\ and\
  \citenamefont {Sfakianakis}}]{Adshead:2016iae}%
  \BibitemOpen
  \bibfield  {author} {\bibinfo {author} {\bibfnamefont {Peter}\ \bibnamefont
  {Adshead}}, \bibinfo {author} {\bibfnamefont {John~T.}\ \bibnamefont
  {Giblin}}, \bibinfo {author} {\bibfnamefont {Timothy~R.}\ \bibnamefont
  {Scully}}, \ and\ \bibinfo {author} {\bibfnamefont {Evangelos~I.}\
  \bibnamefont {Sfakianakis}},\ }\bibfield  {title} {\enquote {\bibinfo {title}
  {{Magnetogenesis from axion inflation}},}\ }\href {\doibase
  10.1088/1475-7516/2016/10/039} {\bibfield  {journal} {\bibinfo  {journal}
  {JCAP}\ }\textbf {\bibinfo {volume} {10}},\ \bibinfo {pages} {039} (\bibinfo
  {year} {2016})},\ \Eprint {http://arxiv.org/abs/1606.08474} {arXiv:1606.08474
  [astro-ph.CO]} \BibitemShut {NoStop}%
\bibitem [{\citenamefont {Anber}\ and\ \citenamefont
  {Sorbo}(2006)}]{Anber:2006xt}%
  \BibitemOpen
  \bibfield  {author} {\bibinfo {author} {\bibfnamefont {Mohamed~M.}\
  \bibnamefont {Anber}}\ and\ \bibinfo {author} {\bibfnamefont {Lorenzo}\
  \bibnamefont {Sorbo}},\ }\bibfield  {title} {\enquote {\bibinfo {title}
  {N-flationary magnetic fields},}\ }\href {\doibase
  10.1088/1475-7516/2006/10/018} {\bibfield  {journal} {\bibinfo  {journal}
  {JCAP}\ }\textbf {\bibinfo {volume} {10}},\ \bibinfo {pages} {018} (\bibinfo
  {year} {2006})},\ \Eprint {http://arxiv.org/abs/astro-ph/0606534}
  {arXiv:astro-ph/0606534 [astro-ph]} \BibitemShut {NoStop}%
\bibitem [{\citenamefont {Kamada}\ and\ \citenamefont
  {Long}(2016{\natexlab{a}})}]{Kamada:2017sxh}%
  \BibitemOpen
  \bibfield  {author} {\bibinfo {author} {\bibfnamefont {Kohei}\ \bibnamefont
  {Kamada}}\ and\ \bibinfo {author} {\bibfnamefont {Andrew~J.}\ \bibnamefont
  {Long}},\ }\bibfield  {title} {\enquote {\bibinfo {title} {Axion production
  from primordial magnetic fields},}\ }\href {\doibase
  10.1103/PhysRevD.94.063501} {\bibfield  {journal} {\bibinfo  {journal} {Phys.
  Rev. D}\ }\textbf {\bibinfo {volume} {94}},\ \bibinfo {pages} {063501}
  (\bibinfo {year} {2016}{\natexlab{a}})},\ \Eprint
  {http://arxiv.org/abs/1606.08891} {arXiv:1606.08891 [hep-ph]} \BibitemShut
  {NoStop}%
\bibitem [{\citenamefont {Joyce}\ and\ \citenamefont
  {Shaposhnikov}(1997{\natexlab{a}})}]{Joyce:1997uy}%
  \BibitemOpen
  \bibfield  {author} {\bibinfo {author} {\bibfnamefont {Michael}\ \bibnamefont
  {Joyce}}\ and\ \bibinfo {author} {\bibfnamefont {Mikhail~E.}\ \bibnamefont
  {Shaposhnikov}},\ }\bibfield  {title} {\enquote {\bibinfo {title} {Primordial
  magnetic fields, right-handed electrons, and the abelian anomaly},}\ }\href
  {\doibase 10.1103/PhysRevLett.79.1193} {\bibfield  {journal} {\bibinfo
  {journal} {Phys. Rev. Lett.}\ }\textbf {\bibinfo {volume} {79}},\ \bibinfo
  {pages} {1193--1196} (\bibinfo {year} {1997}{\natexlab{a}})},\ \Eprint
  {http://arxiv.org/abs/astro-ph/9703005} {arXiv:astro-ph/9703005 [astro-ph]}
  \BibitemShut {NoStop}%
\bibitem [{\citenamefont {Giovannini}\ and\ \citenamefont
  {Shaposhnikov}(1998)}]{Giovannini:1997gp}%
  \BibitemOpen
  \bibfield  {author} {\bibinfo {author} {\bibfnamefont {Massimo}\ \bibnamefont
  {Giovannini}}\ and\ \bibinfo {author} {\bibfnamefont {Mikhail~E.}\
  \bibnamefont {Shaposhnikov}},\ }\bibfield  {title} {\enquote {\bibinfo
  {title} {Primordial hypermagnetic fields and triangle anomaly},}\ }\href
  {\doibase 10.1103/PhysRevD.57.2186} {\bibfield  {journal} {\bibinfo
  {journal} {Phys. Rev. D}\ }\textbf {\bibinfo {volume} {57}},\ \bibinfo
  {pages} {2186--2206} (\bibinfo {year} {1998})},\ \Eprint
  {http://arxiv.org/abs/hep-ph/9710234} {arXiv:hep-ph/9710234 [hep-ph]}
  \BibitemShut {NoStop}%
\bibitem [{\citenamefont {Dvornikov}\ and\ \citenamefont
  {Semikoz}(2020)}]{Dvornikov:2021jpk}%
  \BibitemOpen
  \bibfield  {author} {\bibinfo {author} {\bibfnamefont {Maxim}\ \bibnamefont
  {Dvornikov}}\ and\ \bibinfo {author} {\bibfnamefont {Victor~B.}\ \bibnamefont
  {Semikoz}},\ }\bibfield  {title} {\enquote {\bibinfo {title} {Evolution of
  the primordial magnetic fields in the hot universe: chiral matter and
  turbulence effects},}\ }\href {\doibase 10.1016/j.physrep.2020.08.002}
  {\bibfield  {journal} {\bibinfo  {journal} {Phys. Rept.}\ }\textbf {\bibinfo
  {volume} {887}},\ \bibinfo {pages} {1--83} (\bibinfo {year} {2020})},\
  \Eprint {http://arxiv.org/abs/2005.00272} {arXiv:2005.00272 [hep-ph]}
  \BibitemShut {NoStop}%
\bibitem [{\citenamefont {Durrer}\ and\ \citenamefont
  {Neronov}(2013{\natexlab{a}})}]{Durrer:2013pga}%
  \BibitemOpen
  \bibfield  {author} {\bibinfo {author} {\bibfnamefont {Ruth}\ \bibnamefont
  {Durrer}}\ and\ \bibinfo {author} {\bibfnamefont {Andrii}\ \bibnamefont
  {Neronov}},\ }\bibfield  {title} {\enquote {\bibinfo {title} {{Cosmological
  Magnetic Fields: Their Generation, Evolution and Observation}},}\ }\href
  {\doibase 10.1007/s00159-013-0062-7} {\bibfield  {journal} {\bibinfo
  {journal} {Astron. Astrophys. Rev.}\ }\textbf {\bibinfo {volume} {21}},\
  \bibinfo {pages} {62} (\bibinfo {year} {2013}{\natexlab{a}})},\ \Eprint
  {http://arxiv.org/abs/1303.7121} {arXiv:1303.7121 [astro-ph.CO]} \BibitemShut
  {NoStop}%
\bibitem [{\citenamefont {Sakharov}(1967)}]{Sakharov1967dj}%
  \BibitemOpen
  \bibfield  {author} {\bibinfo {author} {\bibfnamefont {A.~D.}\ \bibnamefont
  {Sakharov}},\ }\bibfield  {title} {\enquote {\bibinfo {title} {{Violation of
  CP Invariance, C asymmetry, and baryon asymmetry of the universe}},}\ }\href
  {\doibase 10.1070/PU1991v034n05ABEH002497} {\bibfield  {journal} {\bibinfo
  {journal} {Pisma Zh. Eksp. Teor. Fiz.}\ }\textbf {\bibinfo {volume} {5}},\
  \bibinfo {pages} {32--35} (\bibinfo {year} {1967})}\BibitemShut {NoStop}%
\bibitem [{\citenamefont {Morrissey}\ and\ \citenamefont
  {Ramsey-Musolf}(2012)}]{Morrissey2012db}%
  \BibitemOpen
  \bibfield  {author} {\bibinfo {author} {\bibfnamefont {David~E.}\
  \bibnamefont {Morrissey}}\ and\ \bibinfo {author} {\bibfnamefont
  {Michael~J.}\ \bibnamefont {Ramsey-Musolf}},\ }\bibfield  {title} {\enquote
  {\bibinfo {title} {{Electroweak baryogenesis}},}\ }\href {\doibase
  10.1088/1367-2630/14/12/125003} {\bibfield  {journal} {\bibinfo  {journal}
  {New J. Phys.}\ }\textbf {\bibinfo {volume} {14}},\ \bibinfo {pages} {125003}
  (\bibinfo {year} {2012})},\ \Eprint {http://arxiv.org/abs/1206.2942}
  {arXiv:1206.2942 [hep-ph]} \BibitemShut {NoStop}%
\bibitem [{\citenamefont {Boyarsky}\ \emph {et~al.}(2012)\citenamefont
  {Boyarsky}, \citenamefont {Frohlich},\ and\ \citenamefont
  {Ruchayskiy}}]{Boyarsky:2011uy}%
  \BibitemOpen
  \bibfield  {author} {\bibinfo {author} {\bibfnamefont {Alexey}\ \bibnamefont
  {Boyarsky}}, \bibinfo {author} {\bibfnamefont {Jurg}\ \bibnamefont
  {Frohlich}}, \ and\ \bibinfo {author} {\bibfnamefont {Oleg}\ \bibnamefont
  {Ruchayskiy}},\ }\bibfield  {title} {\enquote {\bibinfo {title}
  {{Self-consistent evolution of magnetic fields and chiral asymmetry in the
  early Universe}},}\ }\href {\doibase 10.1103/PhysRevLett.108.031301}
  {\bibfield  {journal} {\bibinfo  {journal} {Phys. Rev. Lett.}\ }\textbf
  {\bibinfo {volume} {108}},\ \bibinfo {pages} {031301} (\bibinfo {year}
  {2012})},\ \Eprint {http://arxiv.org/abs/1109.3350} {arXiv:1109.3350
  [astro-ph.CO]} \BibitemShut {NoStop}%
\bibitem [{\citenamefont {Vilenkin}(1980)}]{Vilenkin:1980fu}%
  \BibitemOpen
  \bibfield  {author} {\bibinfo {author} {\bibfnamefont {A.}~\bibnamefont
  {Vilenkin}},\ }\bibfield  {title} {\enquote {\bibinfo {title} {Equilibrium
  parity violating current in a magnetic field},}\ }\href {\doibase
  10.1103/PhysRevD.22.3080} {\bibfield  {journal} {\bibinfo  {journal} {Phys.
  Rev. D}\ }\textbf {\bibinfo {volume} {22}},\ \bibinfo {pages} {3080--3084}
  (\bibinfo {year} {1980})}\BibitemShut {NoStop}%
\bibitem [{\citenamefont {Fukushima}\ \emph {et~al.}(2008)\citenamefont
  {Fukushima}, \citenamefont {Kharzeev},\ and\ \citenamefont
  {Warringa}}]{PhysRevD.78.074033}%
  \BibitemOpen
  \bibfield  {author} {\bibinfo {author} {\bibfnamefont {Kenji}\ \bibnamefont
  {Fukushima}}, \bibinfo {author} {\bibfnamefont {Dmitri~E.}\ \bibnamefont
  {Kharzeev}}, \ and\ \bibinfo {author} {\bibfnamefont {Harmen~J.}\
  \bibnamefont {Warringa}},\ }\bibfield  {title} {\enquote {\bibinfo {title}
  {Chiral magnetic effect},}\ }\href {\doibase 10.1103/PhysRevD.78.074033}
  {\bibfield  {journal} {\bibinfo  {journal} {Phys. Rev. D}\ }\textbf {\bibinfo
  {volume} {78}},\ \bibinfo {pages} {074033} (\bibinfo {year}
  {2008})}\BibitemShut {NoStop}%
\bibitem [{\citenamefont {Akamatsu}\ and\ \citenamefont
  {Yamamoto}(2013)}]{Akamatsu:2013pjd}%
  \BibitemOpen
  \bibfield  {author} {\bibinfo {author} {\bibfnamefont {Yukinao}\ \bibnamefont
  {Akamatsu}}\ and\ \bibinfo {author} {\bibfnamefont {Naoki}\ \bibnamefont
  {Yamamoto}},\ }\bibfield  {title} {\enquote {\bibinfo {title} {Chiral plasma
  instabilities},}\ }\href {\doibase 10.1103/PhysRevLett.111.052002} {\bibfield
   {journal} {\bibinfo  {journal} {Phys. Rev. Lett.}\ }\textbf {\bibinfo
  {volume} {111}},\ \bibinfo {pages} {052002} (\bibinfo {year} {2013})},\
  \Eprint {http://arxiv.org/abs/1302.2125} {arXiv:1302.2125 [nucl-th]}
  \BibitemShut {NoStop}%
\bibitem [{\citenamefont {Rogachevskii}\ \emph {et~al.}(2017)\citenamefont
  {Rogachevskii}, \citenamefont {Ruchayskiy}, \citenamefont {Boyarsky},
  \citenamefont {Fr{\"o}hlich}, \citenamefont {Kleeorin},\ and\ \citenamefont
  {Brandenburg}}]{Rogachevskii:2017uyc}%
  \BibitemOpen
  \bibfield  {author} {\bibinfo {author} {\bibfnamefont {Igor}\ \bibnamefont
  {Rogachevskii}}, \bibinfo {author} {\bibfnamefont {Oleg}\ \bibnamefont
  {Ruchayskiy}}, \bibinfo {author} {\bibfnamefont {Alexey}\ \bibnamefont
  {Boyarsky}}, \bibinfo {author} {\bibfnamefont {J{\"u}rg}\ \bibnamefont
  {Fr{\"o}hlich}}, \bibinfo {author} {\bibfnamefont {Nathan}\ \bibnamefont
  {Kleeorin}}, \ and\ \bibinfo {author} {\bibfnamefont {Axel}\ \bibnamefont
  {Brandenburg}},\ }\bibfield  {title} {\enquote {\bibinfo {title} {Laminar and
  turbulent dynamos in chiral magnetohydrodynamics -- i: Theory},}\ }\href
  {\doibase 10.3847/1538-4357/aa86a6} {\bibfield  {journal} {\bibinfo
  {journal} {Astrophys. J.}\ }\textbf {\bibinfo {volume} {846}},\ \bibinfo
  {pages} {153} (\bibinfo {year} {2017})},\ \Eprint
  {http://arxiv.org/abs/1705.00378} {arXiv:1705.00378 [astro-ph.HE]}
  \BibitemShut {NoStop}%
\bibitem [{\citenamefont {Kamada}\ and\ \citenamefont
  {Long}(2016{\natexlab{b}})}]{Kamada:2016eeb}%
  \BibitemOpen
  \bibfield  {author} {\bibinfo {author} {\bibfnamefont {Kohei}\ \bibnamefont
  {Kamada}}\ and\ \bibinfo {author} {\bibfnamefont {Andrew~J.}\ \bibnamefont
  {Long}},\ }\bibfield  {title} {\enquote {\bibinfo {title} {Baryogenesis from
  decaying magnetic helicity},}\ }\href {\doibase 10.1103/PhysRevD.94.063501}
  {\bibfield  {journal} {\bibinfo  {journal} {Phys. Rev. D}\ }\textbf {\bibinfo
  {volume} {94}},\ \bibinfo {pages} {063501} (\bibinfo {year}
  {2016}{\natexlab{b}})},\ \Eprint {http://arxiv.org/abs/1606.08891}
  {arXiv:1606.08891 [hep-ph]} \BibitemShut {NoStop}%
\bibitem [{\citenamefont {Long}\ \emph {et~al.}(2014)\citenamefont {Long},
  \citenamefont {Sabancilar},\ and\ \citenamefont {Vachaspati}}]{Long:2016lmj}%
  \BibitemOpen
  \bibfield  {author} {\bibinfo {author} {\bibfnamefont {Andrew~J.}\
  \bibnamefont {Long}}, \bibinfo {author} {\bibfnamefont {Eray}\ \bibnamefont
  {Sabancilar}}, \ and\ \bibinfo {author} {\bibfnamefont {Tanmay}\ \bibnamefont
  {Vachaspati}},\ }\bibfield  {title} {\enquote {\bibinfo {title} {Baryogenesis
  from decaying magnetic helicity},}\ }\href {\doibase
  10.1088/1475-7516/2014/02/036} {\bibfield  {journal} {\bibinfo  {journal}
  {JCAP}\ }\textbf {\bibinfo {volume} {02}},\ \bibinfo {pages} {036} (\bibinfo
  {year} {2014})},\ \Eprint {http://arxiv.org/abs/1309.2315} {arXiv:1309.2315
  [astro-ph.CO]} \BibitemShut {NoStop}%
\bibitem [{\citenamefont {Brandenburg}\ \emph {et~al.}(2021)\citenamefont
  {Brandenburg}, \citenamefont {Schober}, \citenamefont {Rogachevskii},\ and\
  \citenamefont {Kahniashvili}}]{Brandenburg:2021lnj}%
  \BibitemOpen
  \bibfield  {author} {\bibinfo {author} {\bibfnamefont {Axel}\ \bibnamefont
  {Brandenburg}}, \bibinfo {author} {\bibfnamefont {Jennifer}\ \bibnamefont
  {Schober}}, \bibinfo {author} {\bibfnamefont {Igor}\ \bibnamefont
  {Rogachevskii}}, \ and\ \bibinfo {author} {\bibfnamefont {Tina}\ \bibnamefont
  {Kahniashvili}},\ }\bibfield  {title} {\enquote {\bibinfo {title} {Relic
  gravitational waves from the chiral magnetic effect},}\ }\href {\doibase
  10.1103/PhysRevD.104.043513} {\bibfield  {journal} {\bibinfo  {journal}
  {Phys. Rev. D}\ }\textbf {\bibinfo {volume} {104}},\ \bibinfo {pages}
  {043513} (\bibinfo {year} {2021})},\ \Eprint
  {http://arxiv.org/abs/2103.01140} {arXiv:2103.01140 [astro-ph.CO]}
  \BibitemShut {NoStop}%
\bibitem [{\citenamefont {{Planck Collaboration}}(2016)}]{Planck:2015zrl}%
  \BibitemOpen
  \bibfield  {author} {\bibinfo {author} {\bibnamefont {{Planck
  Collaboration}}},\ }\bibfield  {title} {\enquote {\bibinfo {title} {Planck
  2015 results. xix. constraints on primordial magnetic fields},}\ }\href
  {\doibase 10.1051/0004-6361/201525821} {\bibfield  {journal} {\bibinfo
  {journal} {Astron. Astrophys.}\ }\textbf {\bibinfo {volume} {594}},\ \bibinfo
  {pages} {A19} (\bibinfo {year} {2016})},\ \Eprint
  {http://arxiv.org/abs/1502.01594} {arXiv:1502.01594 [astro-ph.CO]}
  \BibitemShut {NoStop}%
\bibitem [{\citenamefont {Paoletti}\ and\ \citenamefont
  {Finelli}(2013)}]{Paoletti:2022pfi}%
  \BibitemOpen
  \bibfield  {author} {\bibinfo {author} {\bibfnamefont {Daniela}\ \bibnamefont
  {Paoletti}}\ and\ \bibinfo {author} {\bibfnamefont {Fabio}\ \bibnamefont
  {Finelli}},\ }\bibfield  {title} {\enquote {\bibinfo {title} {Constraints on
  primordial magnetic fields from their impact on the ionization history of the
  universe},}\ }\href {\doibase 10.1016/j.physletb.2013.08.001} {\bibfield
  {journal} {\bibinfo  {journal} {Phys. Lett. B}\ }\textbf {\bibinfo {volume}
  {726}},\ \bibinfo {pages} {45--54} (\bibinfo {year} {2013})},\ \Eprint
  {http://arxiv.org/abs/1305.7248} {arXiv:1305.7248 [astro-ph.CO]} \BibitemShut
  {NoStop}%
\bibitem [{\citenamefont {Lasky}\ \emph {et~al.}(2016)\citenamefont {Lasky}
  \emph {et~al.}}]{Lasky:2015lej}%
  \BibitemOpen
  \bibfield  {author} {\bibinfo {author} {\bibfnamefont {Paul~D.}\ \bibnamefont
  {Lasky}} \emph {et~al.},\ }\bibfield  {title} {\enquote {\bibinfo {title}
  {Gravitational-wave cosmology across 29 decades in frequency},}\ }\href
  {\doibase 10.1103/PhysRevX.6.011035} {\bibfield  {journal} {\bibinfo
  {journal} {Phys. Rev. X}\ }\textbf {\bibinfo {volume} {6}},\ \bibinfo {pages}
  {011035} (\bibinfo {year} {2016})},\ \Eprint
  {http://arxiv.org/abs/1511.05994} {arXiv:1511.05994 [astro-ph.CO]}
  \BibitemShut {NoStop}%
\bibitem [{\citenamefont {Caprini}\ and\ \citenamefont
  {Figueroa}(2018)}]{Caprini:2018mtu}%
  \BibitemOpen
  \bibfield  {author} {\bibinfo {author} {\bibfnamefont {Chiara}\ \bibnamefont
  {Caprini}}\ and\ \bibinfo {author} {\bibfnamefont {Daniel~G.}\ \bibnamefont
  {Figueroa}},\ }\bibfield  {title} {\enquote {\bibinfo {title} {Cosmological
  backgrounds of gravitational waves},}\ }\href {\doibase
  10.1088/1361-6382/aac608} {\bibfield  {journal} {\bibinfo  {journal} {Class.
  Quant. Grav.}\ }\textbf {\bibinfo {volume} {35}},\ \bibinfo {pages} {163001}
  (\bibinfo {year} {2018})},\ \Eprint {http://arxiv.org/abs/1801.04268}
  {arXiv:1801.04268 [astro-ph.CO]} \BibitemShut {NoStop}%
\bibitem [{\citenamefont {Du}\ \emph {et~al.}(2018)\citenamefont {Du} \emph
  {et~al.}}]{ADMX2018gho}%
  \BibitemOpen
  \bibfield  {author} {\bibinfo {author} {\bibfnamefont {N.}~\bibnamefont {Du}}
  \emph {et~al.} (\bibinfo {collaboration} {ADMX}),\ }\bibfield  {title}
  {\enquote {\bibinfo {title} {{A Search for Invisible Axion Dark Matter with
  the Axion Dark Matter Experiment}},}\ }\href {\doibase
  10.1103/PhysRevLett.120.151301} {\bibfield  {journal} {\bibinfo  {journal}
  {Phys. Rev. Lett.}\ }\textbf {\bibinfo {volume} {120}},\ \bibinfo {pages}
  {151301} (\bibinfo {year} {2018})},\ \Eprint
  {http://arxiv.org/abs/1804.05750} {arXiv:1804.05750 [hep-ex]} \BibitemShut
  {NoStop}%
\bibitem [{\citenamefont {{MADMAX Collaboration}}(2022)}]{MADMAX:2022urh}%
  \BibitemOpen
  \bibfield  {author} {\bibinfo {author} {\bibnamefont {{MADMAX
  Collaboration}}},\ }\bibfield  {title} {\enquote {\bibinfo {title} {A new
  experimental approach to probe qcd axion dark matter in the mass range above
  40 $\mu$ev},}\ }\href {\doibase 10.1140/epjc/s10052-021-09882-z} {\bibfield
  {journal} {\bibinfo  {journal} {Eur. Phys. J. C}\ }\textbf {\bibinfo {volume}
  {82}},\ \bibinfo {pages} {105} (\bibinfo {year} {2022})},\ \Eprint
  {http://arxiv.org/abs/2111.02386} {arXiv:2111.02386 [hep-ex]} \BibitemShut
  {NoStop}%
\bibitem [{\citenamefont {{DMRadio Collaboration}}(2022)}]{DMRadio:2022pkf}%
  \BibitemOpen
  \bibfield  {author} {\bibinfo {author} {\bibnamefont {{DMRadio
  Collaboration}}},\ }\bibfield  {title} {\enquote {\bibinfo {title} {Search
  for qcd axion dark matter with dmradio-m$^3$},}\ }\href@noop {} {\bibfield
  {journal} {\bibinfo  {journal} {arXiv e-prints}\ } (\bibinfo {year}
  {2022})},\ \Eprint {http://arxiv.org/abs/2210.01770} {arXiv:2210.01770
  [hep-ex]} \BibitemShut {NoStop}%
\bibitem [{\citenamefont {Anastassopoulos}\ \emph {et~al.}(2017)\citenamefont
  {Anastassopoulos} \emph {et~al.}}]{CAST2017uph}%
  \BibitemOpen
  \bibfield  {author} {\bibinfo {author} {\bibfnamefont {V.}~\bibnamefont
  {Anastassopoulos}} \emph {et~al.} (\bibinfo {collaboration} {CAST}),\
  }\bibfield  {title} {\enquote {\bibinfo {title} {{New CAST Limit on the
  Axion-Photon Interaction}},}\ }\href {\doibase 10.1038/nphys4109} {\bibfield
  {journal} {\bibinfo  {journal} {Nature Phys.}\ }\textbf {\bibinfo {volume}
  {13}},\ \bibinfo {pages} {584--590} (\bibinfo {year} {2017})},\ \Eprint
  {http://arxiv.org/abs/1705.02290} {arXiv:1705.02290 [hep-ex]} \BibitemShut
  {NoStop}%
\bibitem [{\citenamefont {{IAXO Collaboration}}(2019)}]{IAXO:2019mpb}%
  \BibitemOpen
  \bibfield  {author} {\bibinfo {author} {\bibnamefont {{IAXO
  Collaboration}}},\ }\bibfield  {title} {\enquote {\bibinfo {title} {The
  international axion observatory iaxo. letter of intent to the cern sps
  committee},}\ }\href {\doibase 10.1088/1748-0221/14/11/T11002} {\bibfield
  {journal} {\bibinfo  {journal} {JINST}\ }\textbf {\bibinfo {volume} {14}},\
  \bibinfo {pages} {T11002} (\bibinfo {year} {2019})},\ \Eprint
  {http://arxiv.org/abs/1904.09155} {arXiv:1904.09155 [physics.ins-det]}
  \BibitemShut {NoStop}%
\bibitem [{\citenamefont {Abbott}\ \emph {et~al.}(2016)\citenamefont {Abbott}
  \emph {et~al.}}]{LIGOScientific:2016aoc}%
  \BibitemOpen
  \bibfield  {author} {\bibinfo {author} {\bibfnamefont {B.~P.}\ \bibnamefont
  {Abbott}} \emph {et~al.} (\bibinfo {collaboration} {LIGO Scientific,
  Virgo}),\ }\bibfield  {title} {\enquote {\bibinfo {title} {{Observation of
  Gravitational Waves from a Binary Black Hole Merger}},}\ }\href {\doibase
  10.1103/PhysRevLett.116.061102} {\bibfield  {journal} {\bibinfo  {journal}
  {Phys. Rev. Lett.}\ }\textbf {\bibinfo {volume} {116}},\ \bibinfo {pages}
  {061102} (\bibinfo {year} {2016})},\ \Eprint
  {http://arxiv.org/abs/1602.03837} {arXiv:1602.03837 [gr-qc]} \BibitemShut
  {NoStop}%
\bibitem [{\citenamefont {Caprini}\ \emph {et~al.}(2020)\citenamefont {Caprini}
  \emph {et~al.}}]{Caprini:2019egz}%
  \BibitemOpen
  \bibfield  {author} {\bibinfo {author} {\bibfnamefont {Chiara}\ \bibnamefont
  {Caprini}} \emph {et~al.},\ }\bibfield  {title} {\enquote {\bibinfo {title}
  {{Detecting gravitational waves from cosmological phase transitions with
  LISA: an update}},}\ }\href {\doibase 10.1088/1475-7516/2020/03/024}
  {\bibfield  {journal} {\bibinfo  {journal} {JCAP}\ }\textbf {\bibinfo
  {volume} {03}},\ \bibinfo {pages} {024} (\bibinfo {year} {2020})},\ \Eprint
  {http://arxiv.org/abs/1910.13125} {arXiv:1910.13125 [astro-ph.CO]}
  \BibitemShut {NoStop}%
\bibitem [{\citenamefont {Turner}\ and\ \citenamefont
  {Widrow}(1988)}]{Turner1987bw}%
  \BibitemOpen
  \bibfield  {author} {\bibinfo {author} {\bibfnamefont {Michael~S.}\
  \bibnamefont {Turner}}\ and\ \bibinfo {author} {\bibfnamefont {Lawrence~M.}\
  \bibnamefont {Widrow}},\ }\bibfield  {title} {\enquote {\bibinfo {title}
  {{Inflation Produced, Large Scale Magnetic Fields}},}\ }\href {\doibase
  10.1103/PhysRevD.37.2743} {\bibfield  {journal} {\bibinfo  {journal} {Phys.
  Rev. D}\ }\textbf {\bibinfo {volume} {37}},\ \bibinfo {pages} {2743}
  (\bibinfo {year} {1988})}\BibitemShut {NoStop}%
\bibitem [{\citenamefont {Co}\ \emph {et~al.}(2020{\natexlab{b}})\citenamefont
  {Co}, \citenamefont {Hall},\ and\ \citenamefont {Harigaya}}]{Co2019jts}%
  \BibitemOpen
  \bibfield  {author} {\bibinfo {author} {\bibfnamefont {Raymond~T.}\
  \bibnamefont {Co}}, \bibinfo {author} {\bibfnamefont {Lawrence~J.}\
  \bibnamefont {Hall}}, \ and\ \bibinfo {author} {\bibfnamefont {Keisuke}\
  \bibnamefont {Harigaya}},\ }\bibfield  {title} {\enquote {\bibinfo {title}
  {{Axion Kinetic Misalignment Mechanism}},}\ }\href {\doibase
  10.1103/PhysRevLett.124.251802} {\bibfield  {journal} {\bibinfo  {journal}
  {Phys. Rev. Lett.}\ }\textbf {\bibinfo {volume} {124}},\ \bibinfo {pages}
  {251802} (\bibinfo {year} {2020}{\natexlab{b}})},\ \Eprint
  {http://arxiv.org/abs/1910.14152} {arXiv:1910.14152 [hep-ph]} \BibitemShut
  {NoStop}%
\bibitem [{\citenamefont {Chang}\ and\ \citenamefont
  {Cui}(2020{\natexlab{b}})}]{Chang2019tvx}%
  \BibitemOpen
  \bibfield  {author} {\bibinfo {author} {\bibfnamefont {Chia-Feng}\
  \bibnamefont {Chang}}\ and\ \bibinfo {author} {\bibfnamefont {Yanou}\
  \bibnamefont {Cui}},\ }\bibfield  {title} {\enquote {\bibinfo {title} {{New
  Perspectives on Axion Misalignment Mechanism}},}\ }\href {\doibase
  10.1103/PhysRevD.102.015003} {\bibfield  {journal} {\bibinfo  {journal}
  {Phys. Rev. D}\ }\textbf {\bibinfo {volume} {102}},\ \bibinfo {pages}
  {015003} (\bibinfo {year} {2020}{\natexlab{b}})},\ \Eprint
  {http://arxiv.org/abs/1911.11885} {arXiv:1911.11885 [hep-ph]} \BibitemShut
  {NoStop}%
\bibitem [{\citenamefont {Subramanian}(2016)}]{Subramanian2015lua}%
  \BibitemOpen
  \bibfield  {author} {\bibinfo {author} {\bibfnamefont {Kandaswamy}\
  \bibnamefont {Subramanian}},\ }\bibfield  {title} {\enquote {\bibinfo {title}
  {{The origin, evolution and signatures of primordial magnetic fields}},}\
  }\href {\doibase 10.1088/0034-4885/79/7/076901} {\bibfield  {journal}
  {\bibinfo  {journal} {Rept. Prog. Phys.}\ }\textbf {\bibinfo {volume} {79}},\
  \bibinfo {pages} {076901} (\bibinfo {year} {2016})},\ \Eprint
  {http://arxiv.org/abs/1504.02311} {arXiv:1504.02311 [astro-ph.CO]}
  \BibitemShut {NoStop}%
\bibitem [{\citenamefont {Tajima}\ \emph {et~al.}(1992)\citenamefont {Tajima},
  \citenamefont {Cable}, \citenamefont {Shibata},\ and\ \citenamefont
  {Kulsrud}}]{Tajima}%
  \BibitemOpen
  \bibfield  {author} {\bibinfo {author} {\bibfnamefont {T.}~\bibnamefont
  {Tajima}}, \bibinfo {author} {\bibfnamefont {Sam}\ \bibnamefont {Cable}},
  \bibinfo {author} {\bibfnamefont {Kazunari}\ \bibnamefont {Shibata}}, \ and\
  \bibinfo {author} {\bibfnamefont {R.}~\bibnamefont {Kulsrud}},\ }\bibfield
  {title} {\enquote {\bibinfo {title} {On the origin of cosmological magnetic
  fields},}\ }\href {\doibase 10.1086/171281} {\bibfield  {journal} {\bibinfo
  {journal} {The Astrophysical Journal}\ }\textbf {\bibinfo {volume} {390}},\
  \bibinfo {pages} {309} (\bibinfo {year} {1992})}\BibitemShut {NoStop}%
\bibitem [{\citenamefont {Grasso}\ and\ \citenamefont
  {Rubinstein}(2001)}]{Grasso2000wj}%
  \BibitemOpen
  \bibfield  {author} {\bibinfo {author} {\bibfnamefont {Dario}\ \bibnamefont
  {Grasso}}\ and\ \bibinfo {author} {\bibfnamefont {Hector~R.}\ \bibnamefont
  {Rubinstein}},\ }\bibfield  {title} {\enquote {\bibinfo {title} {{Magnetic
  fields in the early universe}},}\ }\href {\doibase
  10.1016/S0370-1573(00)00110-1} {\bibfield  {journal} {\bibinfo  {journal}
  {Phys. Rept.}\ }\textbf {\bibinfo {volume} {348}},\ \bibinfo {pages}
  {163--266} (\bibinfo {year} {2001})},\ \Eprint
  {http://arxiv.org/abs/astro-ph/0009061} {arXiv:astro-ph/0009061} \BibitemShut
  {NoStop}%
\bibitem [{\citenamefont {Vachaspati}(1991)}]{Vachaspati1991nm}%
  \BibitemOpen
  \bibfield  {author} {\bibinfo {author} {\bibfnamefont {T.}~\bibnamefont
  {Vachaspati}},\ }\bibfield  {title} {\enquote {\bibinfo {title} {{Magnetic
  fields from cosmological phase transitions}},}\ }\href {\doibase
  10.1016/0370-2693(91)90051-Q} {\bibfield  {journal} {\bibinfo  {journal}
  {Phys. Lett. B}\ }\textbf {\bibinfo {volume} {265}},\ \bibinfo {pages}
  {258--261} (\bibinfo {year} {1991})}\BibitemShut {NoStop}%
\bibitem [{\citenamefont {Cornwall}(1997)}]{Cornwall1997ms}%
  \BibitemOpen
  \bibfield  {author} {\bibinfo {author} {\bibfnamefont {John~M.}\ \bibnamefont
  {Cornwall}},\ }\bibfield  {title} {\enquote {\bibinfo {title} {{Speculations
  on primordial magnetic helicity}},}\ }\href {\doibase
  10.1103/PhysRevD.56.6146} {\bibfield  {journal} {\bibinfo  {journal} {Phys.
  Rev. D}\ }\textbf {\bibinfo {volume} {56}},\ \bibinfo {pages} {6146--6154}
  (\bibinfo {year} {1997})},\ \Eprint {http://arxiv.org/abs/hep-th/9704022}
  {arXiv:hep-th/9704022} \BibitemShut {NoStop}%
\bibitem [{\citenamefont {Jedamzik}\ \emph
  {et~al.}(1998{\natexlab{a}})\citenamefont {Jedamzik}, \citenamefont
  {Katalini\ifmmode~\acute{c}\else \'{c}\fi{}},\ and\ \citenamefont
  {Olinto}}]{PhysRevD.57.3264}%
  \BibitemOpen
  \bibfield  {author} {\bibinfo {author} {\bibfnamefont {Karsten}\ \bibnamefont
  {Jedamzik}}, \bibinfo {author} {\bibfnamefont {Vi\ifmmode
  \check{s}\else~\v{s}\fi{}nja}\ \bibnamefont {Katalini\ifmmode~\acute{c}\else
  \'{c}\fi{}}}, \ and\ \bibinfo {author} {\bibfnamefont {Angela~V.}\
  \bibnamefont {Olinto}},\ }\bibfield  {title} {\enquote {\bibinfo {title}
  {Damping of cosmic magnetic fields},}\ }\href {\doibase
  10.1103/PhysRevD.57.3264} {\bibfield  {journal} {\bibinfo  {journal} {Phys.
  Rev. D}\ }\textbf {\bibinfo {volume} {57}},\ \bibinfo {pages} {3264--3284}
  (\bibinfo {year} {1998}{\natexlab{a}})}\BibitemShut {NoStop}%
\bibitem [{\citenamefont {Durrer}\ and\ \citenamefont
  {Neronov}(2013{\natexlab{b}})}]{Durrer2013pga}%
  \BibitemOpen
  \bibfield  {author} {\bibinfo {author} {\bibfnamefont {Ruth}\ \bibnamefont
  {Durrer}}\ and\ \bibinfo {author} {\bibfnamefont {Andrii}\ \bibnamefont
  {Neronov}},\ }\bibfield  {title} {\enquote {\bibinfo {title} {{Cosmological
  Magnetic Fields: Their Generation, Evolution and Observation}},}\ }\href
  {\doibase 10.1007/s00159-013-0062-7} {\bibfield  {journal} {\bibinfo
  {journal} {Astron. Astrophys. Rev.}\ }\textbf {\bibinfo {volume} {21}},\
  \bibinfo {pages} {62} (\bibinfo {year} {2013}{\natexlab{b}})},\ \Eprint
  {http://arxiv.org/abs/1303.7121} {arXiv:1303.7121 [astro-ph.CO]} \BibitemShut
  {NoStop}%
\bibitem [{\citenamefont {Widrow}(2002)}]{Widrow2002ud}%
  \BibitemOpen
  \bibfield  {author} {\bibinfo {author} {\bibfnamefont {Lawrence~M.}\
  \bibnamefont {Widrow}},\ }\bibfield  {title} {\enquote {\bibinfo {title}
  {{Origin of galactic and extragalactic magnetic fields}},}\ }\href {\doibase
  10.1103/RevModPhys.74.775} {\bibfield  {journal} {\bibinfo  {journal} {Rev.
  Mod. Phys.}\ }\textbf {\bibinfo {volume} {74}},\ \bibinfo {pages} {775--823}
  (\bibinfo {year} {2002})},\ \Eprint {http://arxiv.org/abs/astro-ph/0207240}
  {arXiv:astro-ph/0207240} \BibitemShut {NoStop}%
\bibitem [{\citenamefont {Brandenburg}\ and\ \citenamefont
  {Subramanian}(2004)}]{Brandenburg2004AstrophysicalMF}%
  \BibitemOpen
  \bibfield  {author} {\bibinfo {author} {\bibfnamefont {Axel}\ \bibnamefont
  {Brandenburg}}\ and\ \bibinfo {author} {\bibfnamefont {Kandaswamy}\
  \bibnamefont {Subramanian}},\ }\bibfield  {title} {\enquote {\bibinfo {title}
  {Astrophysical magnetic fields and nonlinear dynamo theory},}\ }\href
  {https://api.semanticscholar.org/CorpusID:119518712} {\bibfield  {journal}
  {\bibinfo  {journal} {Physics Reports}\ }\textbf {\bibinfo {volume} {417}},\
  \bibinfo {pages} {1--209} (\bibinfo {year} {2004})}\BibitemShut {NoStop}%
\bibitem [{\citenamefont {Banerjee}\ and\ \citenamefont
  {Jedamzik}(2004)}]{Banerjee2004df}%
  \BibitemOpen
  \bibfield  {author} {\bibinfo {author} {\bibfnamefont {Robi}\ \bibnamefont
  {Banerjee}}\ and\ \bibinfo {author} {\bibfnamefont {Karsten}\ \bibnamefont
  {Jedamzik}},\ }\bibfield  {title} {\enquote {\bibinfo {title} {{The Evolution
  of cosmic magnetic fields: From the very early universe, to recombination, to
  the present}},}\ }\href {\doibase 10.1103/PhysRevD.70.123003} {\bibfield
  {journal} {\bibinfo  {journal} {Phys. Rev. D}\ }\textbf {\bibinfo {volume}
  {70}},\ \bibinfo {pages} {123003} (\bibinfo {year} {2004})},\ \Eprint
  {http://arxiv.org/abs/astro-ph/0410032} {arXiv:astro-ph/0410032} \BibitemShut
  {NoStop}%
\bibitem [{\citenamefont {Brandenburg}\ and\ \citenamefont
  {Kahniashvili}(2017)}]{Brandenburg2016odr}%
  \BibitemOpen
  \bibfield  {author} {\bibinfo {author} {\bibfnamefont {Axel}\ \bibnamefont
  {Brandenburg}}\ and\ \bibinfo {author} {\bibfnamefont {Tina}\ \bibnamefont
  {Kahniashvili}},\ }\bibfield  {title} {\enquote {\bibinfo {title} {{Classes
  of hydrodynamic and magnetohydrodynamic turbulent decay}},}\ }\href {\doibase
  10.1103/PhysRevLett.118.055102} {\bibfield  {journal} {\bibinfo  {journal}
  {Phys. Rev. Lett.}\ }\textbf {\bibinfo {volume} {118}},\ \bibinfo {pages}
  {055102} (\bibinfo {year} {2017})},\ \Eprint
  {http://arxiv.org/abs/1607.01360} {arXiv:1607.01360 [physics.flu-dyn]}
  \BibitemShut {NoStop}%
\bibitem [{\citenamefont {Jedamzik}\ \emph
  {et~al.}(1998{\natexlab{b}})\citenamefont {Jedamzik}, \citenamefont
  {Katalinic},\ and\ \citenamefont {Olinto}}]{Jedamzik1996wp}%
  \BibitemOpen
  \bibfield  {author} {\bibinfo {author} {\bibfnamefont {Karsten}\ \bibnamefont
  {Jedamzik}}, \bibinfo {author} {\bibfnamefont {Visnja}\ \bibnamefont
  {Katalinic}}, \ and\ \bibinfo {author} {\bibfnamefont {Angela~V.}\
  \bibnamefont {Olinto}},\ }\bibfield  {title} {\enquote {\bibinfo {title}
  {{Damping of cosmic magnetic fields}},}\ }\href {\doibase
  10.1103/PhysRevD.57.3264} {\bibfield  {journal} {\bibinfo  {journal} {Phys.
  Rev. D}\ }\textbf {\bibinfo {volume} {57}},\ \bibinfo {pages} {3264--3284}
  (\bibinfo {year} {1998}{\natexlab{b}})},\ \Eprint
  {http://arxiv.org/abs/astro-ph/9606080} {arXiv:astro-ph/9606080} \BibitemShut
  {NoStop}%
\bibitem [{\citenamefont {Subramanian}\ and\ \citenamefont
  {Barrow}(1998)}]{Subramanian1997gi}%
  \BibitemOpen
  \bibfield  {author} {\bibinfo {author} {\bibfnamefont {Kandaswamy}\
  \bibnamefont {Subramanian}}\ and\ \bibinfo {author} {\bibfnamefont {John~D.}\
  \bibnamefont {Barrow}},\ }\bibfield  {title} {\enquote {\bibinfo {title}
  {{Magnetohydrodynamics in the early universe and the damping of noninear
  Alfven waves}},}\ }\href {\doibase 10.1103/PhysRevD.58.083502} {\bibfield
  {journal} {\bibinfo  {journal} {Phys. Rev. D}\ }\textbf {\bibinfo {volume}
  {58}},\ \bibinfo {pages} {083502} (\bibinfo {year} {1998})},\ \Eprint
  {http://arxiv.org/abs/astro-ph/9712083} {arXiv:astro-ph/9712083} \BibitemShut
  {NoStop}%
\bibitem [{\citenamefont {Schwarz}\ and\ \citenamefont
  {Stuke}(2009)}]{Schwarz2009ii}%
  \BibitemOpen
  \bibfield  {author} {\bibinfo {author} {\bibfnamefont {Dominik~J.}\
  \bibnamefont {Schwarz}}\ and\ \bibinfo {author} {\bibfnamefont {Maik}\
  \bibnamefont {Stuke}},\ }\bibfield  {title} {\enquote {\bibinfo {title}
  {{Lepton asymmetry and the cosmic QCD transition}},}\ }\href {\doibase
  10.1088/1475-7516/2009/11/025} {\bibfield  {journal} {\bibinfo  {journal}
  {JCAP}\ }\textbf {\bibinfo {volume} {11}},\ \bibinfo {pages} {025} (\bibinfo
  {year} {2009})},\ \bibinfo {note} {[Erratum: JCAP 10, E01 (2010)]},\ \Eprint
  {http://arxiv.org/abs/0906.3434} {arXiv:0906.3434 [hep-ph]} \BibitemShut
  {NoStop}%
\bibitem [{\citenamefont {Kahniashvili}\ \emph {et~al.}(2010)\citenamefont
  {Kahniashvili}, \citenamefont {Tevzadze}, \citenamefont {Sethi},
  \citenamefont {Pandey},\ and\ \citenamefont {Ratra}}]{Kahniashvili2010wm}%
  \BibitemOpen
  \bibfield  {author} {\bibinfo {author} {\bibfnamefont {Tina}\ \bibnamefont
  {Kahniashvili}}, \bibinfo {author} {\bibfnamefont {Alexander~G.}\
  \bibnamefont {Tevzadze}}, \bibinfo {author} {\bibfnamefont {Shiv~K.}\
  \bibnamefont {Sethi}}, \bibinfo {author} {\bibfnamefont {Kanhaiya}\
  \bibnamefont {Pandey}}, \ and\ \bibinfo {author} {\bibfnamefont {Bharat}\
  \bibnamefont {Ratra}},\ }\bibfield  {title} {\enquote {\bibinfo {title}
  {{Primordial magnetic field limits from cosmological data}},}\ }\href
  {\doibase 10.1103/PhysRevD.82.083005} {\bibfield  {journal} {\bibinfo
  {journal} {Phys. Rev. D}\ }\textbf {\bibinfo {volume} {82}},\ \bibinfo
  {pages} {083005} (\bibinfo {year} {2010})},\ \Eprint
  {http://arxiv.org/abs/1009.2094} {arXiv:1009.2094 [astro-ph.CO]} \BibitemShut
  {NoStop}%
\bibitem [{\citenamefont {Raffelt}\ and\ \citenamefont
  {Stodolsky}(1988)}]{Raffelt1987im}%
  \BibitemOpen
  \bibfield  {author} {\bibinfo {author} {\bibfnamefont {Georg}\ \bibnamefont
  {Raffelt}}\ and\ \bibinfo {author} {\bibfnamefont {Leo}\ \bibnamefont
  {Stodolsky}},\ }\bibfield  {title} {\enquote {\bibinfo {title} {{Mixing of
  the Photon with Low Mass Particles}},}\ }\href {\doibase
  10.1103/PhysRevD.37.1237} {\bibfield  {journal} {\bibinfo  {journal} {Phys.
  Rev. D}\ }\textbf {\bibinfo {volume} {37}},\ \bibinfo {pages} {1237}
  (\bibinfo {year} {1988})}\BibitemShut {NoStop}%
\bibitem [{\citenamefont {Mirizzi}\ \emph {et~al.}(2007)\citenamefont
  {Mirizzi}, \citenamefont {Raffelt},\ and\ \citenamefont
  {Serpico}}]{Mirizzi2007hr}%
  \BibitemOpen
  \bibfield  {author} {\bibinfo {author} {\bibfnamefont {Alessandro}\
  \bibnamefont {Mirizzi}}, \bibinfo {author} {\bibfnamefont {Georg~G.}\
  \bibnamefont {Raffelt}}, \ and\ \bibinfo {author} {\bibfnamefont
  {Pasquale~D.}\ \bibnamefont {Serpico}},\ }\bibfield  {title} {\enquote
  {\bibinfo {title} {{Signatures of Axion-Like Particles in the Spectra of TeV
  Gamma-Ray Sources}},}\ }\href {\doibase 10.1103/PhysRevD.76.023001}
  {\bibfield  {journal} {\bibinfo  {journal} {Phys. Rev. D}\ }\textbf {\bibinfo
  {volume} {76}},\ \bibinfo {pages} {023001} (\bibinfo {year} {2007})},\
  \Eprint {http://arxiv.org/abs/0704.3044} {arXiv:0704.3044 [astro-ph]}
  \BibitemShut {NoStop}%
\bibitem [{\citenamefont {Marsh}(2016{\natexlab{b}})}]{Marsh2015xka}%
  \BibitemOpen
  \bibfield  {author} {\bibinfo {author} {\bibfnamefont {David J.~E.}\
  \bibnamefont {Marsh}},\ }\bibfield  {title} {\enquote {\bibinfo {title}
  {{Axion Cosmology}},}\ }\href {\doibase 10.1016/j.physrep.2016.06.005}
  {\bibfield  {journal} {\bibinfo  {journal} {Phys. Rept.}\ }\textbf {\bibinfo
  {volume} {643}},\ \bibinfo {pages} {1--79} (\bibinfo {year}
  {2016}{\natexlab{b}})},\ \Eprint {http://arxiv.org/abs/1510.07633}
  {arXiv:1510.07633 [astro-ph.CO]} \BibitemShut {NoStop}%
\bibitem [{\citenamefont {Domcke}\ \emph
  {et~al.}(2019{\natexlab{a}})\citenamefont {Domcke}, \citenamefont {von
  Harling}, \citenamefont {Morgante},\ and\ \citenamefont
  {Mukaida}}]{Domcke2019mnd}%
  \BibitemOpen
  \bibfield  {author} {\bibinfo {author} {\bibfnamefont {Valerie}\ \bibnamefont
  {Domcke}}, \bibinfo {author} {\bibfnamefont {Benedict}\ \bibnamefont {von
  Harling}}, \bibinfo {author} {\bibfnamefont {Enrico}\ \bibnamefont
  {Morgante}}, \ and\ \bibinfo {author} {\bibfnamefont {Kyohei}\ \bibnamefont
  {Mukaida}},\ }\bibfield  {title} {\enquote {\bibinfo {title} {{Baryogenesis
  from axion inflation}},}\ }\href {\doibase 10.1088/1475-7516/2019/10/032}
  {\bibfield  {journal} {\bibinfo  {journal} {JCAP}\ }\textbf {\bibinfo
  {volume} {10}},\ \bibinfo {pages} {032} (\bibinfo {year}
  {2019}{\natexlab{a}})},\ \Eprint {http://arxiv.org/abs/1905.13318}
  {arXiv:1905.13318 [hep-ph]} \BibitemShut {NoStop}%
\bibitem [{\citenamefont {Kamada}\ and\ \citenamefont
  {Long}(2016{\natexlab{c}})}]{PhysRevD.94.123509}%
  \BibitemOpen
  \bibfield  {author} {\bibinfo {author} {\bibfnamefont {Kohei}\ \bibnamefont
  {Kamada}}\ and\ \bibinfo {author} {\bibfnamefont {Andrew~J.}\ \bibnamefont
  {Long}},\ }\bibfield  {title} {\enquote {\bibinfo {title} {Evolution of the
  baryon asymmetry through the electroweak crossover in the presence of a
  helical magnetic field},}\ }\href {\doibase 10.1103/PhysRevD.94.123509}
  {\bibfield  {journal} {\bibinfo  {journal} {Phys. Rev. D}\ }\textbf {\bibinfo
  {volume} {94}},\ \bibinfo {pages} {123509} (\bibinfo {year}
  {2016}{\natexlab{c}})}\BibitemShut {NoStop}%
\bibitem [{\citenamefont {Joyce}\ and\ \citenamefont
  {Shaposhnikov}(1997{\natexlab{b}})}]{PhysRevLett.79.1193}%
  \BibitemOpen
  \bibfield  {author} {\bibinfo {author} {\bibfnamefont {M.}~\bibnamefont
  {Joyce}}\ and\ \bibinfo {author} {\bibfnamefont {M.}~\bibnamefont
  {Shaposhnikov}},\ }\bibfield  {title} {\enquote {\bibinfo {title} {Primordial
  magnetic fields, right electrons, and the abelian anomaly},}\ }\href
  {\doibase 10.1103/PhysRevLett.79.1193} {\bibfield  {journal} {\bibinfo
  {journal} {Phys. Rev. Lett.}\ }\textbf {\bibinfo {volume} {79}},\ \bibinfo
  {pages} {1193--1196} (\bibinfo {year} {1997}{\natexlab{b}})}\BibitemShut
  {NoStop}%
\bibitem [{\citenamefont {Kamada}\ and\ \citenamefont
  {Long}(2016{\natexlab{d}})}]{PhysRevD.94.063501}%
  \BibitemOpen
  \bibfield  {author} {\bibinfo {author} {\bibfnamefont {Kohei}\ \bibnamefont
  {Kamada}}\ and\ \bibinfo {author} {\bibfnamefont {Andrew~J.}\ \bibnamefont
  {Long}},\ }\bibfield  {title} {\enquote {\bibinfo {title} {Baryogenesis from
  decaying magnetic helicity},}\ }\href {\doibase 10.1103/PhysRevD.94.063501}
  {\bibfield  {journal} {\bibinfo  {journal} {Phys. Rev. D}\ }\textbf {\bibinfo
  {volume} {94}},\ \bibinfo {pages} {063501} (\bibinfo {year}
  {2016}{\natexlab{d}})}\BibitemShut {NoStop}%
\bibitem [{\citenamefont {Bödeker}\ and\ \citenamefont
  {Schröder}(2019)}]{Bodeker_2019}%
  \BibitemOpen
  \bibfield  {author} {\bibinfo {author} {\bibfnamefont {Dietrich}\
  \bibnamefont {Bödeker}}\ and\ \bibinfo {author} {\bibfnamefont {Dennis}\
  \bibnamefont {Schröder}},\ }\bibfield  {title} {\enquote {\bibinfo {title}
  {Equilibration of right-handed electrons},}\ }\href {\doibase
  10.1088/1475-7516/2019/05/010} {\bibfield  {journal} {\bibinfo  {journal}
  {Journal of Cosmology and Astroparticle Physics}\ }\textbf {\bibinfo {volume}
  {2019}},\ \bibinfo {pages} {010} (\bibinfo {year} {2019})}\BibitemShut
  {NoStop}%
\bibitem [{\citenamefont {Kolb}\ and\ \citenamefont {Turner}(1990)}]{1990The}%
  \BibitemOpen
  \bibfield  {author} {\bibinfo {author} {\bibfnamefont {Edward~W}\
  \bibnamefont {Kolb}}\ and\ \bibinfo {author} {\bibfnamefont
  {Michael~Stanley}\ \bibnamefont {Turner}},\ }\bibfield  {title} {\enquote
  {\bibinfo {title} {The early universe},}\ }\href@noop {} {\bibfield
  {journal} {\bibinfo  {journal} {Front. Phys., Vol. 69,}\ }\textbf {\bibinfo
  {volume} {294}},\ \bibinfo {pages} {521--526} (\bibinfo {year}
  {1990})}\BibitemShut {NoStop}%
\bibitem [{\citenamefont {Di~Luzio}\ \emph
  {et~al.}(2020{\natexlab{b}})\citenamefont {Di~Luzio}, \citenamefont
  {Giannotti}, \citenamefont {Nardi},\ and\ \citenamefont
  {Visinelli}}]{DiLuzio2020wdo}%
  \BibitemOpen
  \bibfield  {author} {\bibinfo {author} {\bibfnamefont {Luca}\ \bibnamefont
  {Di~Luzio}}, \bibinfo {author} {\bibfnamefont {Maurizio}\ \bibnamefont
  {Giannotti}}, \bibinfo {author} {\bibfnamefont {Enrico}\ \bibnamefont
  {Nardi}}, \ and\ \bibinfo {author} {\bibfnamefont {Luca}\ \bibnamefont
  {Visinelli}},\ }\bibfield  {title} {\enquote {\bibinfo {title} {{The
  landscape of QCD axion models}},}\ }\href {\doibase
  10.1016/j.physrep.2020.06.002} {\bibfield  {journal} {\bibinfo  {journal}
  {Phys. Rept.}\ }\textbf {\bibinfo {volume} {870}},\ \bibinfo {pages} {1--117}
  (\bibinfo {year} {2020}{\natexlab{b}})},\ \Eprint
  {http://arxiv.org/abs/2003.01100} {arXiv:2003.01100 [hep-ph]} \BibitemShut
  {NoStop}%
\bibitem [{\citenamefont {Domcke}\ \emph
  {et~al.}(2019{\natexlab{b}})\citenamefont {Domcke}, \citenamefont {von
  Harling}, \citenamefont {Morgante},\ and\ \citenamefont
  {Mukaida}}]{Domcke:2019mnd}%
  \BibitemOpen
  \bibfield  {author} {\bibinfo {author} {\bibfnamefont {Valerie}\ \bibnamefont
  {Domcke}}, \bibinfo {author} {\bibfnamefont {Benedict}\ \bibnamefont {von
  Harling}}, \bibinfo {author} {\bibfnamefont {Enrico}\ \bibnamefont
  {Morgante}}, \ and\ \bibinfo {author} {\bibfnamefont {Kyohei}\ \bibnamefont
  {Mukaida}},\ }\bibfield  {title} {\enquote {\bibinfo {title} {{Baryogenesis
  from axion inflation}},}\ }\href {\doibase 10.1088/1475-7516/2019/10/032}
  {\bibfield  {journal} {\bibinfo  {journal} {JCAP}\ }\textbf {\bibinfo
  {volume} {10}},\ \bibinfo {pages} {032} (\bibinfo {year}
  {2019}{\natexlab{b}})},\ \Eprint {http://arxiv.org/abs/1905.13318}
  {arXiv:1905.13318 [hep-ph]} \BibitemShut {NoStop}%
\bibitem [{\citenamefont {Aghanim}\ \emph {et~al.}(2020)\citenamefont {Aghanim}
  \emph {et~al.}}]{Planck2018nkj}%
  \BibitemOpen
  \bibfield  {author} {\bibinfo {author} {\bibfnamefont {N.}~\bibnamefont
  {Aghanim}} \emph {et~al.} (\bibinfo {collaboration} {Planck}),\ }\bibfield
  {title} {\enquote {\bibinfo {title} {{Planck 2018 results. I. Overview and
  the cosmological legacy of Planck}},}\ }\href {\doibase
  10.1051/0004-6361/201833880} {\bibfield  {journal} {\bibinfo  {journal}
  {Astron. Astrophys.}\ }\textbf {\bibinfo {volume} {641}},\ \bibinfo {pages}
  {A1} (\bibinfo {year} {2020})},\ \Eprint {http://arxiv.org/abs/1807.06205}
  {arXiv:1807.06205 [astro-ph.CO]} \BibitemShut {NoStop}%
\end{thebibliography}%

\end{document}